\documentclass[aps, prl, reprint, amsmath,amssymb,nofootinbib, superscriptaddress, showpacs]{revtex4-1}
\usepackage{graphicx}
\usepackage{dcolumn}
\usepackage{bm}
\usepackage{color}
\usepackage{amsmath}
\usepackage{amssymb}
\usepackage{graphicx}
\usepackage{url}
\usepackage{lineno}
\usepackage[colorlinks=false]{hyperref}
\usepackage[textsize=footnotesize]{todonotes}
\usepackage[todos]{CMImacros}
\usepackage{ulem}

\newcommand{\angstrom}{\text{\normalfont\AA}}
\usepackage{color}
\definecolor{oldtxtcolor}{rgb}{0.00, 0.0, 0.5}
\definecolor{newtxtcolor}{rgb}{0.00, 0.3867, 0.00}
\definecolor{newtxtcolor}{rgb}{0.00, 0.0, 0}
\definecolor{oldtxtcolor}{rgb}{1.00, 0.0, 0.00}

\def\verX{12}
\def\verO{1}
\def\verN{2}
\def\verON{12}

\ifx\verX\verO
 \newcommand { \oldtxt }[1] {{\color{oldtxtcolor}{#1}}}
 \newcommand { \newtxt }[1] {}
\fi
\ifx\verX\verN
 \newcommand { \oldtxt }[1] {}
 \newcommand { \newtxt }[1] {{\color{newtxtcolor}{#1}}}
\fi
\ifx\verX\verON
 \newcommand { \oldtxt }[1] {{\color{oldtxtcolor}{#1}}}
 \newcommand { \newtxt }[1] {{\color{newtxtcolor}{#1}}}
\fi

\begin{document}
\title{%
Selective quantum Zeno effect of ultracold atom-molecule scattering in dynamic magnetic fields}
\author{Hanwei Yang*}
\affiliation{State Key Laboratory for Mesoscopic Physics and Collaborative Innovation Center of Quantum Matter, School of Physics, Peking University, Beijing 10087, China}
\author{Zunqi Li*} 
\email{These authors contributed equally to this work.}
\affiliation{State Key Laboratory for Mesoscopic Physics and Collaborative Innovation Center of Quantum Matter, School of Physics, Peking University, Beijing 10087, China}
\author{Songbin Zhang} 
\affiliation{Department of Physics, Shaanxi Normal University, Xi'an 710119, People's Republic of China}
\author{Lushuai Cao} 
\affiliation{Wuhan National Laboratory for Optoelectronics and School of Physics, Huazhong University of Science and Technology, Wuhan 430074, China}
\author{John Bohn}
\affiliation{JILA, University of Colorado, Boulder, Colorado 80309, USA}
\author{Shutao Zhang}
\affiliation{State Key Laboratory for Mesoscopic Physics and Collaborative Innovation Center of Quantum Matter, School of Physics, Peking University, Beijing 10087, China}
\author{Haitan Xu} 
\email{xuht@sustech.edu.cn}
\affiliation{School of Physical Sciences, University of Science and Technology of China, Hefei 230026, China}
\affiliation{Shenzhen Institute for Quantum Science and Engineering, Southern University of Science and Technology, Shenzhen 518055, China}
\author{Gaoren Wang} 
\email{gaoren.wang@dlut.edu.cn}
\affiliation{Department of Physics, Dalian Institute of Technology, Dalian, China}
\author{Zheng Li}
\email{zheng.li@pku.edu.cn}
\affiliation{State Key Laboratory for Mesoscopic Physics and Collaborative Innovation Center of Quantum Matter, School of Physics, Peking University, Beijing 10087, China}

\begin{abstract}
We demonstrated that final states of ultracold scattering between atom and molecule can be selectively produced using dynamic magnetic fields of multiple frequencies.
The mechanism of the dynamic magnetic field control is based on a generalized quantum Zeno effect for the selected scattering channels.
In particular, we use an atom-molecule spin flip scattering to show that the transition to the selected final spin projection of the molecule in the inelastic scattering can be suppressed by dynamic modulation of coupling between the Floquet engineered initial and final states.
\end{abstract}

\maketitle

The ancient Zeno's arrow paradox claims that a flying arrow seems not moving when being instantly observed.
This concept is generalized to the quantum Zeno effect (QZE) which states that one can freeze the evolution of a quantum system by frequent measurement.
\newtxt{In this context, Kofman and Kurizki (KK) proposed dynamical control of quantum mechanical decay based on continuous modulation of the coupling to an ancillary system~\cite{PhysRevLett.87.270405}, which is proved to be equivalent to the concept of "frequent observation" of the system since a "measurement" is nothing but an interaction with an external system playing the role of apparatus~\cite{Facchi02:080401,Facchi04:032314}, and the decay is suppressed owing to an coupling modulation.} 
Such effects have been broadly applied in quantum optics, quantum computation and quantum communication~\cite{Tang18:203902,Longhi08:143602,Barontini15:1317,Bretheau15:776,Silva12:080501}.

Adjusting the microscopic collision via external fields has always been ultimate target in molecular dynamics~\cite{lasercontrol,PhysRevLett.96.123202,prl2015}. 
This goal has stimulated the development of quantum control of molecular processes~\cite{bo1}, which have achieved amazing results in unimolecular chemical reactions~\cite{naturefesh,naturestatis}. 
Attaining control over molecular collisions with simple physical pictures, however, has proven to be a much bigger challenge due to the complexity of the molecular interaction in the rotational, translational and spin degrees of freedom of the system~\cite{B910118G}. 
%
Recent theory and experiments~\cite{PhysRevA.65.052712,PhysRevA.75.022702,Tscherbul_2009,PhysRevLett.98.213001,PhysRevLett.103.183201} demonstrate that inelastic collisions in an ultracold gas of molecular mixture can be effectively tuned by applying a static magnetic field. 
Furthermore, enhancing the rate of specifically chosen reaction channels stimulated the development of quantum control schemes such as optimal control and coherent control~\cite{bo1,tscherbul21:153403}.

\newtxt{Despite of studies on tuning the scattering with static magnetic field, the effects of time dependent magnetic fields on the product state distributions and branching ratios for the different reaction channels remain largely unknown. 
For the transition between bound states, it was found that the transition rate can be reduced by using a single frequency modulation of the external field~\cite{singlezeno}. 
In addition, Smith et al. proposed that a single frequency magnetic field can enhance the pairwise interactions resonantly for the elastic scattering with initial and final states being both unbound~\cite{smith}. 
However, for the more general case of inelastic scattering, the simple physical picture of quantum control with an external field of single frequency could break down due to the complexity of the final state spectrum.}

\begin{figure*}[htp]
    \centering
    \includegraphics[width=14cm]{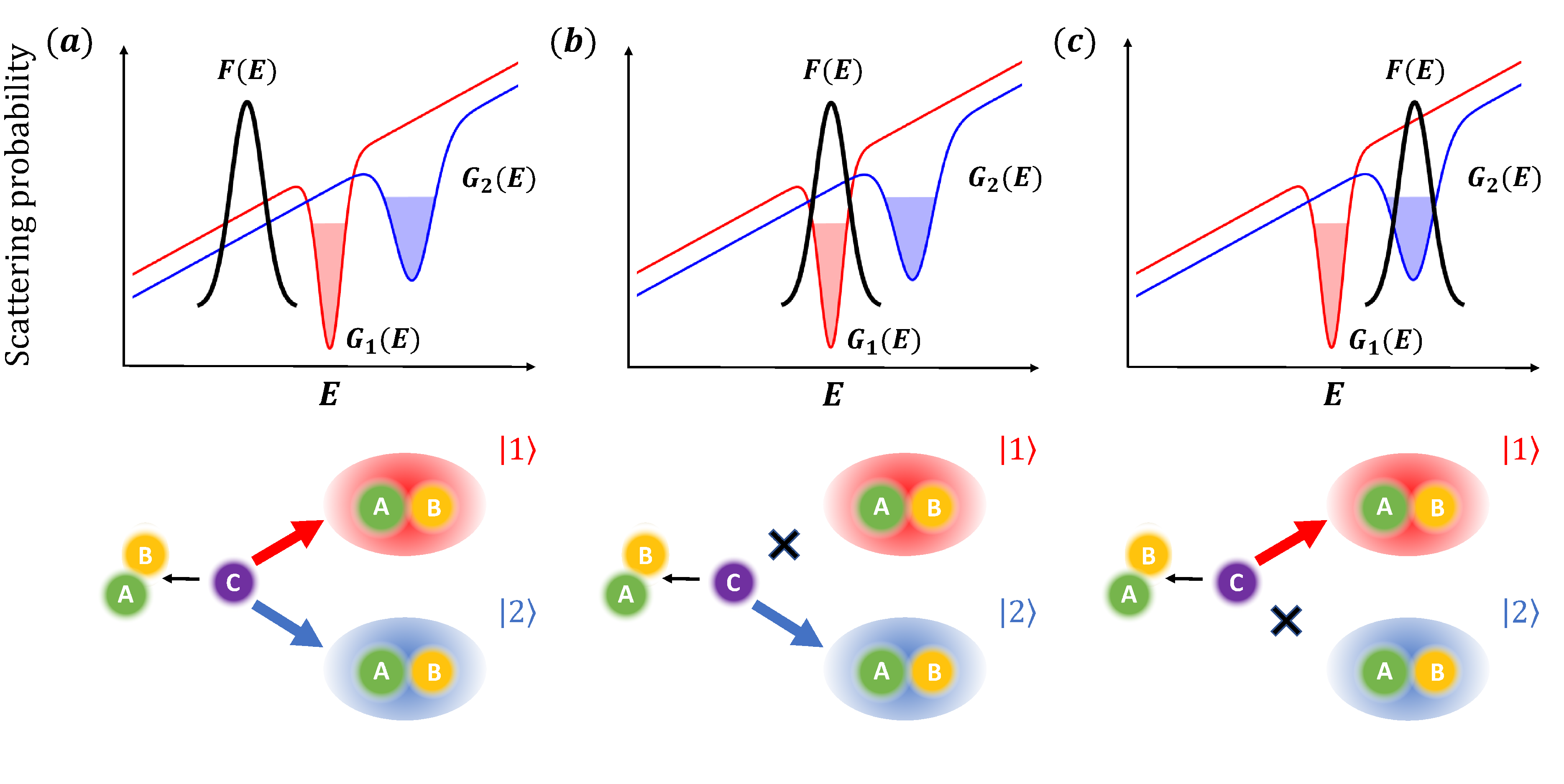}
    \caption{ \label{fig:flc}%
    Schematic diagram of the selective quantum Zeno effect in the collision between a molecule AB and an atom C.
    The red and blue curves represent the transition probabilities $G_f(E)$ from initial state of the molecule to two different final states \ket{f} ($f=1,2$) after the collision as a function of final state energy $E$, which have minima at different energies.
    The spectral modulation function $F(E)$ reflects the temporal variation of eigenenergies of the collision complex induced by external field in the Zeno dynamics, its overlap with the scattering probability $G_f(E)$ determines the cross section for transition to final state $\ket{f}$, i.e. $\sigma_f\propto \int F(E)G_f(E)dE$. 
    By choosing a proper external field, we can make the distribution of $F(E)$ coincide with one of the valleys of $G_f(E)$, and the cross section of the corresponding channel is then suppressed, while other channels are left almost intact.
    In this manner, we can engineer the spectral modulation function to either keep the yields of two final states balanced (a), suppress the yield of final state \ket{1} (red part in the lower panel of (b)) or \ket{2} (blue part in the lower panel (c)).
    }
\end{figure*}
In this Letter, we demonstrate a theory of selective quantum Zeno effect (SQZE) for multi-channel inelastic scattering by a time dependent magnetic pulse train, which is schematically illustrated in Fig.~\ref{fig:flc}.
Our theory permits inverse Floquet engineering of external control fields with clear physical picture, by which we can determine the waveform of magnetic field to selectively suppress one of the product channels.
Numerically, we incorporate multi-frequency Floquet theory into coupled channel calculation to characterize the interaction of the collision complex with the pulsed magnetic field, and show \newtxt{with a concrete example} that the cross sections of different spin-flip channels of the scattering between $\rm{^3He}$ with $\rm{^{17}O_2}$ at ultracold temperatures can be precisely controlled.
This work realizes final state selection in ultracold inelastic scattering, and suggests that a broad range of collision processes are submissive to this dynamic control method, including resonant magneto-association~\cite{PhysRevLett.115.135301,PhysRevLett.85.728} and multi-channel reactive collision~\cite{prl2015}.
%
%

We begin by outlining the multi-frequency Floquet quantum scattering approach used to treat the interaction of collision complex with the magnetic pulse train. 
For the collisions of an atom with oxygen molecule in its ground electronic states $^{3}\Sigma_g^-$ in pulsed magnetic field, we use the Floquet Hamiltonian for the system (in units of $\hbar$)
\begin{equation}
\hat{H}_{\rm{F}}=-\frac{1}{2\mu R}\frac{\partial^2}{\partial R^2}R+\frac{\hat{l} ^2}{2\mu R^2}+\hat{V}(\boldsymbol{R},\boldsymbol{r})+\hat{H}_{\rm{as}}-i\frac{\partial}{\partial t}
\,,
\end{equation}
where $R$ is the atom-molecule separation, $\mu$ is the reduced mass of the collision complex, $\hat{l}$ is the orbital angular momentum of the collision, $r$ is the internuclear distance between atoms in the diatomic molecule, {and $\hat{V}(\boldsymbol{R},\boldsymbol{r})$ is the interaction potential between atom and molecule}~\cite{spinde}. 
The asymptotic Hamiltonian $\hat{H}_{\rm{as}}$ depicts the rotational motion of the isolated molecule and interaction of its electron spin $S$ with the time-dependent magnetic field $B(t)$ through Zeeman effect.
%
%
The total wave function of the collision complex can be expressed in a direct product basis, 
$\Psi = R^{-1}\sum_{\alpha KlM_l}F^{lm_{l}}_{\alpha K}(R)\Phi_{\alpha K}|{lm_{l}}\rangle$,
where $m_l$ denotes the projection of $\hat{l}$ on the magnetic field axis, 
$\Phi_{\alpha K}$ is the eigenstate of the asymptotic Floquet Hamiltonian $\hat{H} _{\rm{asF}}=\hat{H} _{\rm{as}}-i\frac{\partial}{\partial t}$, 
$\alpha$ indicates different channels and $K$ is the index of Floquet eigenstates. 

The magnetic pulse train $B(t)=B_0+\mathcal{B}(t)$ consists of a set of oscillatory magnetic fields of different frequencies. 
Here $B_0$ is the static part and $\mathcal{B}(t)=\sum_{n} {{a_n}}\cos(n\omega_{\mathrm{B}} t)$ is the pulsed field containing multiple Fourier components of frequencies $n\omega_{\mathrm{B}}$. 
We expand the asymptotic Floquet eigenstates $\Phi_{\alpha K}$ in Fourier basis as $\Phi_{\alpha K}=\sum_{NM_N SM_S n}W_{\alpha K,N S J M_J n}|N S J M_J n\rangle$,
where $\hat{N}$ is the rotational angular momentum, $\hat{J}=\hat{N}+\hat{S}$ and $M_J$ is the projection of $J$ on the magnetic field axis, $|n\rangle$ satisfies $\langle t|n\rangle=e^{in\omega_{\mathrm{B}} t}$. 
Since the couplings between states of different $N$s and $J$s are weak compared to the rotational energy, $N$, $J$ and $M_J$ can be considered approximately as good quantum numbers~\cite{PhysRevA.65.052712}, we use $|N S J M_J\rangle$ to label different collision channels. 
The asymptotic Floquet Hamiltonian matrix can be expressed as $\langle \alpha n|H_{\rm{asF}}|\beta m\rangle=H^{n-m}_{\alpha\beta}+n\omega_{\mathrm{B}}\delta_{\alpha\beta}\delta_{nm}$, where $H^{n-m}_{\alpha\beta}$ are Fourier components of $H_{\rm{as}}$, i.e. $\langle \alpha |\hat{H} _{\rm{as}}|\beta \rangle=\sum_{k=-\infty}^{\infty} H_{\alpha \beta}^k e^{ik\omega_B t}$. 
The energy difference between $\Phi_{\alpha K}$ and $\Phi_{\alpha, K+L}$ is $L\omega_{\mathrm{B}}$.
And the total wavefunction is calculated using multi-frequency Floquet coupled channel (MFF-CC) formalism (see SI for details).
%

\begin{figure}
    \centering
    \includegraphics[width=0.48\textwidth]{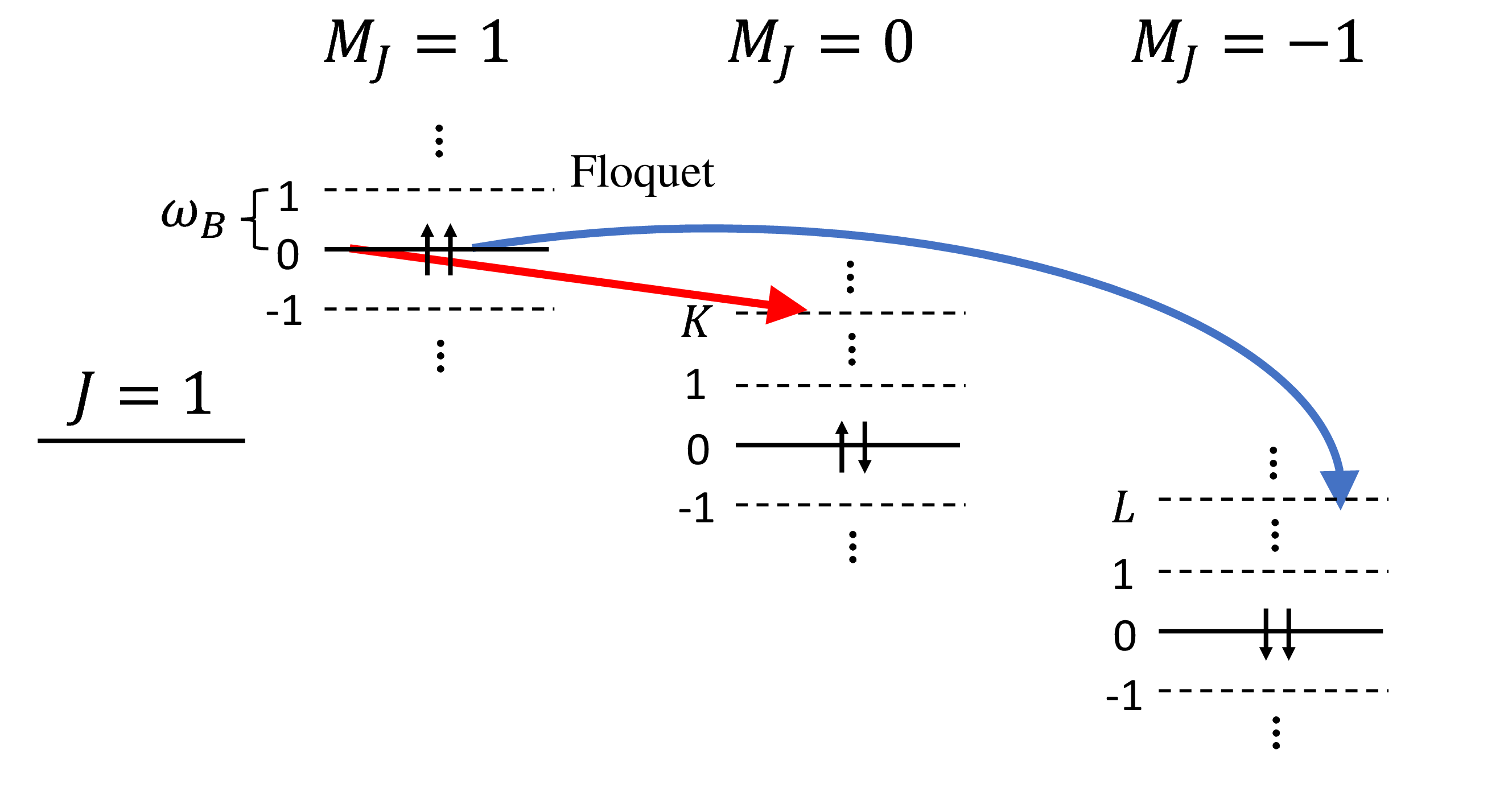}
    \caption{ \label{fig:mech}The mechanism of spin flipping process $| 1,0 \rangle \to | 0,K\rangle$ (red arrows) and $| 1,0\rangle \to |-1,L\rangle$ (blue arrows). The dashed lines indicate the series of Floquet levels with energy spacing of $\omega_{\mathrm{B}}$.}
\end{figure}

We now apply the MFF-CC method to study the effects of pulsed magnetic fields on the spin-flip collision $\mathrm{^3He}\mathrm{+^{17}O_2}(M_J=1)\to\mathrm{^3He}\mathrm{+^{17}O_2}(M_J=0,-1)$ at temperature $\sim 1\mathrm{\mu K}$ with two different channels giving final spin states of $M_J=0$ and $M_J=-1$.
A schematic diagram in Fig.~\ref{fig:mech} demonstrates the mechanism of transitions between spin and Floquet states \newtxt{while the laboratory projection of total angular momentum $M_J+m_l$ is conserved}.
%
%
The channel-resolved cross sections of the collision have minima at distinct \newtxt{external magnetic field} for the two final spin states~\cite{PhysRevA.65.052712}, and provides the possibility to suppress the scattering in the selected channel.
We drop $N,S,J$ since they remain unchanged \newtxt{for the initial and final states of the collision considered in this work}, and abbreviate $|N S J M_J K\rangle$ as $|M_J K\rangle$, and focus on the inelastic collision for the initial spin-up state $|1,0\rangle$, the two spin-flip channels are $|1,0\rangle \to |0,K\rangle$ and $|1,0\rangle \to |-1,L\rangle$. 
We assume the energies of $|M_J=0,0\rangle,|M_J=\pm 1,0\rangle$ to be $e_{0},e_{\pm 1}$. 
For the magnetic field considered in this work, $e_{+1}-e_{0}$ and $e_{0}-e_{-1}$ are equal and proportional to $B_{0}$, we can set $e_{0}=0$ and $e_{\pm 1}=\pm hB_{0}$, where $h=1.2202\times 10^{-4}~\rm{K}/\rm{G}$ for the $\rm{^{17}O_2}$ molecule in $^{3}\Sigma_g^-$ state. 
The kinetic energy after inelastic collision is $E_{\rm{out}}=E_{\rm{in}}+hB_{0}-K\omega_{\mathrm{B}}$ and $E_{\rm{out}}=E_{\rm{in}}+2hB_{0}-L\omega_{\mathrm{B}}$ for the $|1,0\rangle \to |0,K\rangle$ and $|1,0\rangle \to |-1,L\rangle$ channels, respectively. 
\newtxt{It should be noted that the asymptotic energies of states with $M_J=0,\pm1$ in laboratory frame are $0,\pm hB(t)$, thus the total energy before the collision is $E_{\rm{in}}+hB(t)=E_{\rm{in}}+hB_0+h\mathcal{B}(t)$ and it will change to $E_{\rm{out}}+0=E_{\rm{in}}+hB_0-K\omega_B$ or $E_{\rm{ out}}-hB(t)=E_{\rm{in}}+hB_0-h\mathcal{B}(t)-L\omega_B$ for the scattering $|1,0\rangle \to |0,K\rangle$ or $|1,0\rangle \to |-1,L\rangle$, i.e. the inelastic collision produces populations in a series of evenly spaced energy levels with spacing of $\omega_B$}.
%


\newtxt{We still use the asymptotic molecular states $|1,0\rangle$, $|0,K\rangle$ and $|-1,L\rangle$ to label distinct levels whose total energies are $E_{\rm{in}}+hB_0+h\mathcal{B}(t)$, $E_{\rm{in}}+hB_0-K\omega_B$ and $E_{\rm{in}}+hB_0-h\mathcal{B}(t)-L\omega_B$, respectively.}  To reveal the underlying mechanism of the selective magneto-quantum Zeno effect, we model the inelastic scattering as transitions from initial state $|1,0\rangle$ to a series of final states $\{|0,K\rangle\}$ and $\{|-1,L\rangle\}$ of $\rm{^{17}O_2}$ molecule 
with the effective Hamiltonian $\hat{H}(t)=\hat{H}_{0}+\hat{V}+H_{1}(t)$~\cite{PhysRevLett.87.270405}, where
\begin{eqnarray}
\hat{H}_{0}&=& hB_0|1,0\rangle\langle 1,0|+ \sum_{K}(hB_0-K\omega_{\mathrm{B}})|0,K\rangle\langle 0,K|\nonumber\\
&+&\sum_{L}(hB_0-L\omega_{\mathrm{B}})|-1,L\rangle\langle -1,L|
\end{eqnarray}
with $hB_0$, $hB_0-K\omega_{\mathrm{B}}$ and $hB_0-L\omega_{\mathrm{B}}$ being the time-independent energies of $|1,0\rangle$, $|0,K\rangle$ and $|-1,L\rangle$. \newtxt{Here we neglect the constant $E_{\rm{in}}$ in the formula.} And
\begin{eqnarray}
\hat{V}&=&\sum_{K}V_{0}(hB_0-K\omega_{\mathrm{B}})|1,0\rangle\langle 0,K|\nonumber\\
&+&\sum_{L}V_{-1}(2hB_0-L\omega_{\mathrm{B}})|1,0\rangle\langle -1,L|+\mathrm{H.c.}
\end{eqnarray}
stands for the effective coupling of the initial state of molecule $|1,0\rangle$ with the final states $\{|0,K\rangle\},\{|-1,L\rangle\}$, where the coupling strengths $V_0$ and $V_{-1}$ depends on the exit kinetic energy $E_{\rm{out}}$. 
We define the transition probability functions $G_0(E_{\rm{out}})\propto |V_0(E_{\rm{out}})|^2$ and $G_{-1}(E_{\rm{out}})\propto |V_{-1}(E_{\rm{out}})|^2$, \newtxt{which should be understood as the average of coupling and independent of time}.
%
%
The molecule-field interaction
\begin{equation}
\hat{H}_{1}(t)=h\mathcal{B}(t)|1,0\rangle\langle 1,0|-\sum_{L}h\mathcal{B}(t)|-1,L\rangle\langle -1,L|
\end{equation}
depicts the perturbations of molecular energies due to the time dependent magnetic field.
\begin{figure}
    \centering
    \includegraphics[width=7.5cm]{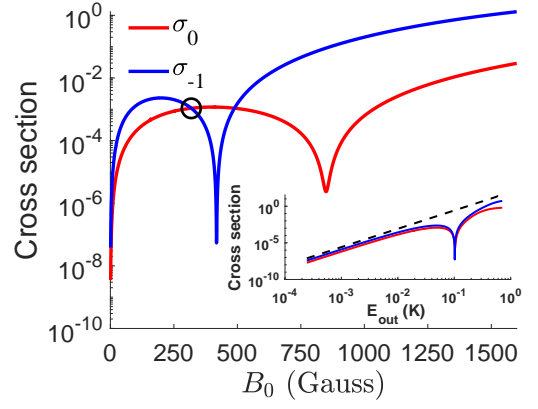}
    \caption{\label{fig:xs} $\rm{^3He}+\rm{^{17}O_2}$ inelastic cross sections~(in $\AA{}^2$) of $M_J=1\to M_J=0$~($\sigma_0$) and $M_J=1\to M_J=-1$~($\sigma_{-1}$) as a function of static magnetic field strength $B_0$ with collision energy $E_{\rm{in}}=1~\rm{\mu K}$. The point marked by the black circle represents the values of $\sigma_0$ and $\sigma_{-1}$ are equal when $B_0=320~\rm{G}$. The inset shows the dependency of $\sigma_0,\sigma_{-1}$ on exit kinetic energy $E_{\rm{out}}$, where the minima of $\sigma_0$ and $\sigma_{-1}$ both  correspond to $E_{\rm{out}}=0.1027~\rm{K}$. The dashed black line shows the $(E_{\rm{out}})^{5/2}$ scaling of the cross section.}
\end{figure}

The wave function of the system at time $t$ can be approximated as
\begin{eqnarray}
\Psi(t)&=&c_1(t)e^{-ihB_0t-i\int_0^t h\mathcal{B}(t')dt'}|1,0\rangle+\nonumber\\
&\sum_K& c_{0,K}(t)e^{-i(hB_0-K\omega_{\mathrm{B}} )t}|0,K\rangle+\nonumber\\
&\sum_L& \!c_{\!-\!1,L}(t)e^{-i(hB_0-L\omega_{\mathrm{B}} )t+i\int_0^t h\mathcal{B}(t')dt'}|-1,L\rangle\,,
\end{eqnarray}
with the initial condition being $c_1(t)=1, c_{0,K}(t)=0, c_{-1,L}(t)=0$.
Because transition to $\ket{1,M}$ with $M\ne 0$ is of higher order, the amplitude $c_{1,M}$ is negligible.
For the spin-flip transition $M_J=1\to M_J=0$, we introduce a modulation function $\epsilon_0(t)=\exp[-i\int_0^t\delta_{1,0}(t')dt']$ and expand it in Fourier series $\epsilon_0(t)=\sum_K \lambda_{0,K} e^{-iK\omega_{\mathrm{B}} t}$, where $\delta_{1,0}(t)=h\mathcal{B}(t)$ is the modulated energies difference between $|1,0\rangle$ and $\{|0,K\rangle\}$.
Similarly we introduce $\epsilon_{-1}(t)=\exp[-i\int_0^t\delta_{1,-1}(t')dt']=\sum_L \lambda_{-1,L} e^{-iL \omega_{\mathrm{B}} t}$ for $M_J=1\to M_J=-1$ transition, where $\delta_{1,-1}(t)=2h\mathcal{B}(t)$.
This temporally modulated system can be treated perturbatively~\cite{PhysRevLett.87.270405,Facchi02:080401,PhysRevLett.97.260402}, which gives the transition rates $\partial_t|c_{0,K}(t)|^2\propto|\lambda_{0,-K}|^2G_0(hB_0-K\omega_{\mathrm{B}})$ and $\partial_t|c_{-1,L}(t)|^2\propto|\lambda_{-1,-L}|^2G_{-1}(2hB_0-L\omega_{\mathrm{B}})$ (see SI for details).
%
%
The cross sections $\sigma_{0,K}$ and $\sigma_{-1,L}$ for the transitions from $|1,0\rangle$ to the final states $|0,K\rangle$ and $|-1,L\rangle$ are proportional to $|\lambda_{0,-K}|^2G_0(hB_0-K\omega_{\mathrm{B}})$ and $|\lambda_{-1,-L}|^2G_{-1}(2hB_0-L\omega_{\mathrm{B}})$ \newtxt{as well}. 
We set up the spectral functions $F_0(E)=\sum_K |\lambda_{0,K}|^2\delta(E-hB_0-K\omega_{\mathrm{B}})$ and $F_{-1}(E)=\sum_L |\lambda_{-1,L}|^2\delta(E-2hB_0-L\omega_{\mathrm{B}})$ which account for the modulation of exit energies $E_{\rm{out}}$.
Let $\sigma_0=\sum_K\sigma_{0,K}$ and $\sigma_{-1}=\sum_L\sigma_{-1,L}$ denote total cross sections of the two spin-flip channels, we have $\sigma_0\propto\int_{-\infty}^{+\infty}F_0(E)G_0(E)dE=\sum_K |\lambda_{0,K}|^2G_0(hB_0+K\omega_{\mathrm{B}})$ and $\sigma_{-1}\propto\int_{-\infty}^{+\infty}F_{-1}(E)G_{-1}(E)dE=\sum_L |\lambda_{-1,L}|^2G_{-1}(2hB_0+L\omega_{\mathrm{B}})$.
%
In particular, when only the static magnetic field $B_0$ is applied, we obtain $\epsilon_0(t)=\epsilon_{-1}(t)=1$ and $\sigma_0\propto G_0(hB_0)$, $\sigma_{-1}\propto G_{-1}(2hB_0)$.
Fig.~\ref{fig:xs} shows the inelastic cross sections as a function of static magnetic field at $E_{\rm{in}}=1~\rm{\mu K}$ with the potential calculated by M$\phi$ller-Plesset perturbation theory~\cite{PhysRevA.65.052712}.
The sharp minima of the cross sections arise from interference between the incident $s$ and emergent $d$ radial wave functions in the inelastic scattering~\cite{PhysRevA.65.052712}.

\begin{figure*}[htp]
    \centering
    \includegraphics[width=7.5cm]{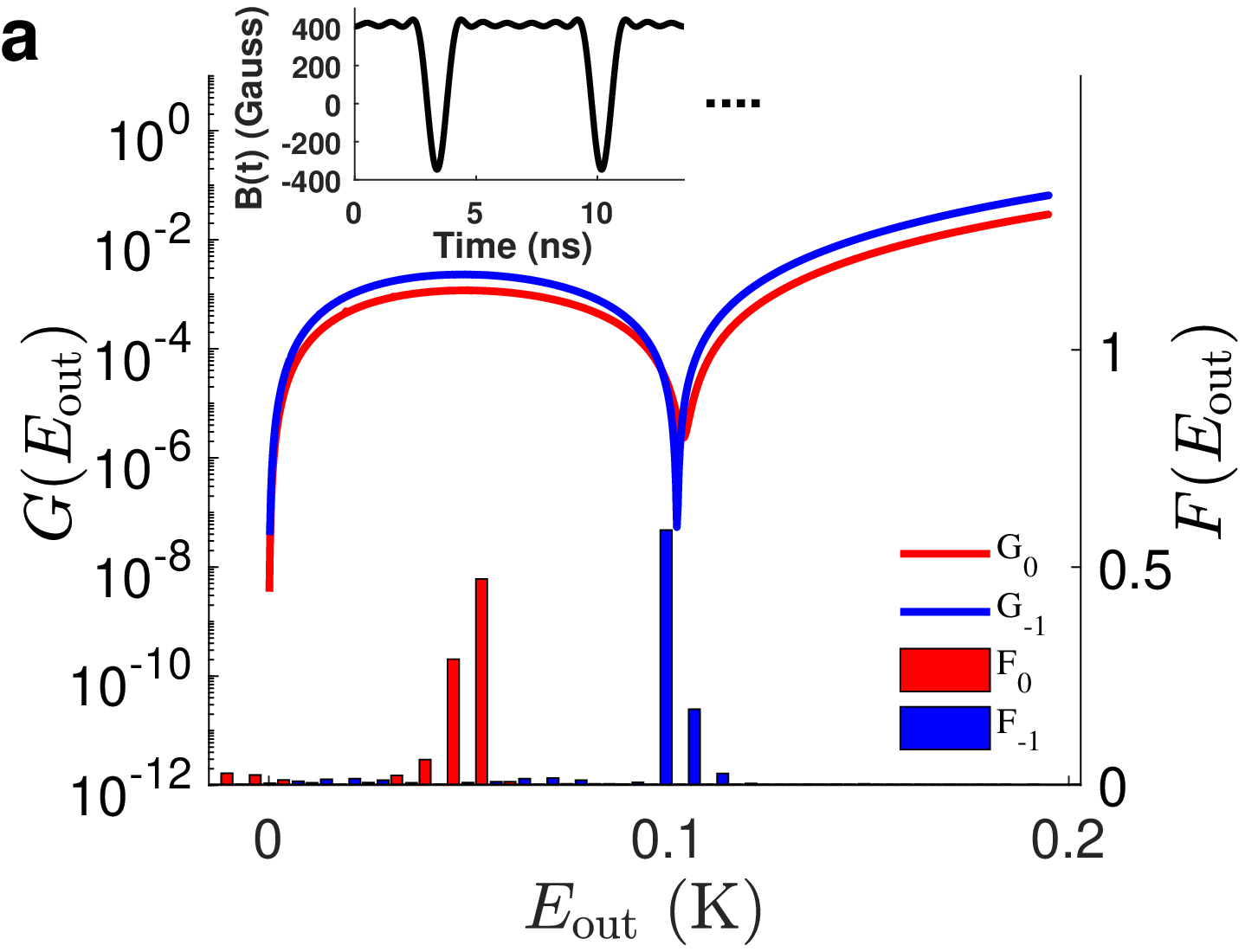}
    \includegraphics[width=7.5cm]{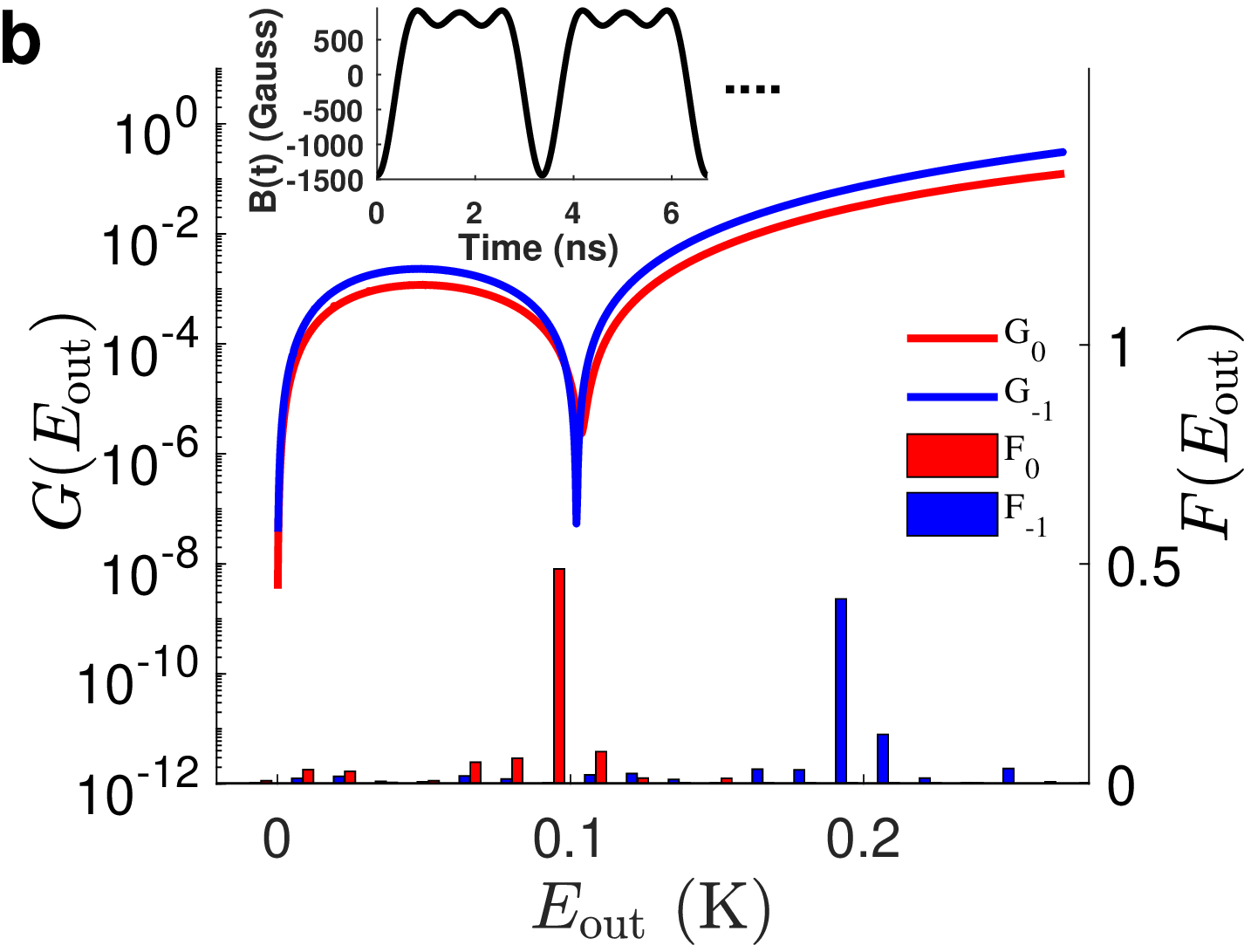}
    \caption{ \label{fig:pr}Transition probabilities as a function of the exit kinetic energy $E_{\rm{out}}$ for the scattering channel $M_J=1\to M_J=0$~($G_0$) and $M_J=1\to M_J=-1$~($G_{-1}$). Notice that the values of $G_0(hB_0),G_{-1}(2hB_0)$ are equal to $\sigma_0,\sigma_{-1}$ respectively when only the static field $B_0$ is applied. In order to select one channel, we choose the time-varying field $B(t)=B_0+\mathcal{B}(t)$, where $B_0=320~G$ and the specific form of $\mathcal{B}(t)$ is given in the main text, to concentrate $F_0(E_{\rm{out}})$ or $F_{-1}(E_{\rm{out}})$ around the minimum. The inset depicts two pulses of the pulse train $B(t)$ in time domain. \textbf{a}, $\mathcal{B}_1(t)=\sum_{n=1}^6 a_n \cos(n\omega_{\mathrm{B}} t)$ and $\sigma_0/\sigma_{-1}=5.16$. \textbf{b}, $\mathcal{B}_2(t)=\sum_{n=1}^3 a_n \cos(n\omega_{\mathrm{B}} t)$ and $\sigma_{-1}/\sigma_{0}=209.42$.}
\end{figure*}

In order to obtain $G_0(E_{\rm{out}})$ and $G_{-1}(E_{\rm{out}})$, we used series of trial magnetic pulses to calculate the Fourier coefficients $\lambda_{0,K}, \lambda_{-1,L}$. 
And by applying the multi-frequency Floquet coupled channel algorithm, we calculated the inelastic cross sections $\sigma_{0,K},\sigma_{-1,L}$ numerically.
Using the relationships $G_0(hB_0-K\omega_{\mathrm{B}})\propto \sigma_{0,K}/|\lambda_{0,-K}|^2$ and $G_{-1}(2hB_0-L\omega_{\mathrm{B}})\propto \sigma_{-1,L}/|\lambda_{-1,-L}|^2$, we retrieve the relative values of $G_0(hB_0-K\omega_{\mathrm{B}})$ and $G_{-1}(2hB_0-L\omega_{\mathrm{B}})$. 
Fig.~\ref{fig:pr} illustrates the calculated transition probabilities $G_f(E_{\mathrm{out}})$ at $E_{\rm{in}}=1~\rm{\mu K}$.
%
%
Within the range of the collision energy and the magnetic field considered here, the wave inelastic scattering is dominated by the $s$ partial wave, and the spin-flip transitions require boosting the angular momentum from $l=0$ to $l=2$. Since the centrifugal barrier of $\sim 0.59$ K for the $d$ partial wave in the exit channel is much larger than $E_{\rm{out}}$, Wigner's threshold law is well reflected by the threshold behavior of transition probabilities, that $G_0$ and $G_{-1}$ are proportional to $(E_{\rm{out}})^{5/2}$~\cite{PhysRevA.65.052712}.

By just tuning the time dependent part of the magnetic field $\mathcal{B}(t)$ and keeping $B_0$ constant, we can dramatically vary the scattering cross sections of different channels and select the final states.
The selection of different scattering channels can be easily implemented by condensing the channel resolved spectral function $F_0(E_{\rm{out}})$ (or $F_{-1}(E_{\rm{out}})$) around the valley of corresponding transition probabilities $G_0(E_{\rm{out}})$ (or $G_{-1}(E_{\rm{out}})$), and the scattering cross section of $M_J=1\to M_J=0$ or $M_J=1\to M_J=-1$ will be effectively suppressed. 
In order to facilitate the comparison, we choose $B_0=320~\rm{G}$ such that the cross sections $\sigma_0, \sigma_{-1}$ are equal.
%
Taking the fundamental frequency $\omega_{\mathrm{B}}$ and amplitude $a_n$ of the time varying part of magnetic field $\mathcal{B}(t)$ as variables and $\int_{-\infty}^{+\infty}F_f(E)G_f(E)dE$ as the objective function, we use genetic algorithm to find the optimal magnetic field to suppress one of the scattering channels. 
To suppress the $M_J=1\to M_J=0$ channel, the optimized field has $\omega_{\mathrm{B}}=2\pi\times 147.46~{\rm{MHz}},a_1=184.40~{\rm{G}},a_2=-165.31~{\rm{G}},a_3=132.02~{\rm{G}},a_4=-95.21~{\rm{G}},a_5=59.59~{\rm{G}},a_6=-30.52~{\rm{G}}$. 
The distributions of channel resolved modulation functions $F_0(E_{\rm{out}})$ and $F_{-1}(E_{\rm{out}})$ are plotted in Fig.~\ref{fig:pr}(a). 
The channel resolved cross sections are $\sigma_{0}=9.7835\times 10^{-4}\rm{\angstrom^2},\sigma_{-1}=1.8959\times 10^{-4}\rm{\angstrom^2}$ and $\sigma_{0}$ is about five times as large as $\sigma_{-1}$. 
To suppress the other channel $M_J=1\to M_J=-1$, we obtained the optimized field with $\omega_{\mathrm{B}}=2\pi\times 297.41~{\rm{MHz}},a_1=-853.72~{\rm{G}},a_2=-595.55~{\rm{G}},a_3=-312.67~{\rm{G}}$, and the distributions of modulation functions are plotted in Fig.~\ref{fig:pr}(b). 
The magnitude of the cross sections are completely reversed, $\sigma_{0}=2.2945\times 10^{-4} \rm{\angstrom^2}$, $\sigma_{-1}=4.8051\times 10^{-2}\rm{\angstrom^2}$, and $\sigma_{-1}$ is two orders of magnitude larger than $\sigma_{0}$. 

\newtxt{Though the collision energy $E_{\rm{in}}$ has a certain Maxwellian distribution, the shapes of $G_0(E_{\rm{out}})$ and $G_{-1}(E_{\rm{out}})$ are almost independent of $E_{\rm{in}}$ for ultracold collision, our control scheme is robust for a thermal distribution of the incident kinetic energy (see explanation in SI).} 
Although the time dependent magnetic field can generate electric field, $\rm{^{17}O_2}$ has no permanent dipole moment, the energy shift (about $ 10~\rm{kHz}$) caused by the Stark effect is much smaller than the Zeeman splitting and Floquet level spacing, hence we only need to consider the effect of magnetic field. 
The method is also applicable to more complex pulse trains consisting of low-frequency magnetic field (about 1~kHz) that can be more easily achieved in experiment. 
%

In conclusion, we have demonstrated a selective quantum Zeno effect that allows flexible tuning of inelastic scattering cross sections by the pulsed magnetic field, and can realize effective selection of scattering channels. 
We have also introduced a theoretical method for solving the quantum scattering problem in the presence of a pulsed external field based on the multi-frequency Floquet approach.
The SQZE mechanism can be directly extended to control broad types of multichannel inelastic and reactive scattering processes with time dependent magnetic, microwave and laser fields of complex temporal structure.
%

The authors are thankful to G. Cha\l{}asi\'nski for helpful discussions. 
This work has been supported by National Natural Science Foundation of China (No. 12174009).

%

\end{document}



\title{Supplementary Information\\
Selective quantum Zeno effect of ultracold atom molecule scattering in dynamic magnetic fields}
\author{Hanwei Yang*}
\affiliation{State Key Laboratory for Mesoscopic Physics and Collaborative Innovation Center of Quantum Matter, School of Physics, Peking University, Beijing 10087, China}
\author{Zunqi Li*} 
\affiliation{State Key Laboratory for Mesoscopic Physics and Collaborative Innovation Center of Quantum Matter, School of Physics, Peking University, Beijing 10087, China}
\author{Songbin Zhang} 
\affiliation{Department of Physics, Shaanxi Normal University, Xi'an 710119, People's Republic of China}
\author{Lushuai Cao} 
\affiliation{Wuhan National Laboratory for Optoelectronics and School of Physics, Huazhong University of Science and Technology, Wuhan 430074, China}
\author{John Bohn}
\affiliation{JILA, University of Colorado, Boulder, Colorado 80309, USA}
\author{Shutao Zhang}
\affiliation{State Key Laboratory for Mesoscopic Physics and Collaborative Innovation Center of Quantum Matter, School of Physics, Peking University, Beijing 10087, China}
\author{Haitan Xu} 
\email{xuht@sustech.edu.cn}
\affiliation{School of Physical Sciences, University of Science and Technology of China, Hefei 230026, China}
\affiliation{Shenzhen Institute for Quantum Science and Engineering, Southern University of Science and Technology, Shenzhen 518055, China}
\author{Gaoren Wang} 
\email{gaoren.wang@dlut.edu.cn}
\affiliation{Department of Physics, Dalian Institute of Technology, Dalian, China}
\author{Zheng Li}
\email{zheng.li@pku.edu.cn}
\affiliation{State Key Laboratory for Mesoscopic Physics and Collaborative Innovation Center of Quantum Matter, School of Physics, Peking University, Beijing 10087, China}

\date{\today}

\maketitle

\begin{quotation}
This supplemental material provides detailed derivation and numerical implementation of our molecular selection method under an time-dependent external magnetic field. The material is organized as follows. In Sec. \uppercase\expandafter{\romannumeral1}, we give a general derivation of the scattering cross section of an inelastic scattering in the presence of external time-dependent magnetic field. In Sec. \uppercase\expandafter{\romannumeral2} , We present a detailed numerical implementation of the scattering cross section of several specific molecular scattering process. Sec. \uppercase\expandafter{\romannumeral3} describes our methods of engineering and regulating the scattering channel by adjusting the spectrum of the external magnetic field.
\end{quotation}

\section{\label{sec:level1}Derivation of the inelastic cross section in the presence of external field via multi-frequency Floquet formalism}
In this section, we shall first review some key results of Floquet theory in A. Then we will derive the Coupled-channel equations and the expression of the scattering wave function. With these preparations, we can then get the explicit form of the scattering cross section of different channels. Finally, in C, we'll compare the process of elastic and inelastic scattering under time-dependent external field, using interaction picture.

\subsection{\label{sec:level2}The Floquet theory}
For a periodic-driven system, we have:
\begin{equation}
    i\frac{\partial}{\partial t}\ket{\Psi(t)}=H(t)\ket{\Psi(t)},\quad H(t+T)=H(t)
\end{equation}
According to Floquet theorem, there're a series of Floquet solutions to this equation with the forms as follows:
\begin{equation}
    \ket{\psi_{\lambda}(t)}=e^{-i\epsilon_{\lambda}t}\ket{\phi_{\lambda}(t)}
\end{equation}
where $\ket{\phi_{\lambda}(t)}$ is a period function in t. The general solution of the Schrodinger equation is the linear combination of these Floquet solutions:
\begin{equation}
    \ket{\Psi(t)}=\sum_{\lambda}a_{\lambda}e^{-i\epsilon_{\lambda}t}\ket{\phi_{\lambda}(t)}
\end{equation}

$\ket{\phi_{\lambda}}$ is subjected to a revised version of Schrodinger equation:
\begin{equation}
    \epsilon_{\lambda}\ket{\phi_{\lambda}}=\left(H-i\frac{\partial}{\partial t}\right)\ket{\phi_{\lambda}}
\end{equation}
Now we define a new Hilbert space, including the time dimension, with a new inner product as:
\begin{equation}
    (\phi|\psi)=\frac{1}{T}\int_{0}^{T}dt\braket{\phi(t)|\psi(t)}
\end{equation}
where $|\phi)$ and $|\psi)$ denote the vectors in the new Hilbert space, corresponding to the time-period state vectors $\ket{\phi}$ and $\ket{\psi}$ respectively.

We expand the Schrodinger equation with a complete set of basis. First we choose a complete set of basis of the original Hilbert space, $\{\ket{\alpha}\}$. Then we expand $|\phi_{\lambda})$ with respect to $e^{in\omega t}\ket{\alpha}$. The expansion coefficients are $\phi_{\alpha \lambda}^{n}=\frac{1}{T}\int_{0}^{T}dt e^{-in\omega t}\braket{\alpha|\phi(t)}$. Then we have:
\begin{equation}
    \epsilon_{\lambda}\phi_{\alpha,\lambda}^{n}=\sum_{\beta,m}H_F^{\alpha,\beta}\phi_{\beta,\lambda}^{m} 
\end{equation}
where 
\begin{eqnarray}
H_F^{\alpha,\beta}&=H_{\alpha,\beta}^{n-m}+n\omega\delta_{\alpha\beta}\delta_{nm}\\
H_{\alpha,\beta}^{n-m}&=\frac{1}{T}\int_{0}^{T}dt \braket{\alpha|H|\beta}e^{i(m-n)\omega t}
\end{eqnarray}
\subsection{\label{sec:level2}Floquet scattering cross section for inelastic scattering}
Consider a two-body scattering process under a time-dependent external field. Without loss of generality, we assume that the Hamiltonian of system consists of two parts:
\begin{equation}
    \hat{H}=\hat{H^0}+\hat{V}(t,\vec{r})
\end{equation}
where $H^0$ is time-independent, and the time average of $V(t)$ is 0.

In most cases, asymptotically, $\hat{V}(t)$ doesn't depend on $r$ and $\theta$(the distance between two molecules and the orientation of their connection), when $r$ approaches infinity. Thus we can separate $\hat{V}(t)$ into two parts, namely the asymptotic part and the interaction part.
\begin{equation}
    \hat{V}(t,\vec{r})=\hat{V}^{as}(t)+\hat{V}^{i}(t,\vec{r})
\end{equation}
where $\hat{V}^{i}(t,\vec{r})\rightarrow 0$ as $r \rightarrow \infty$.

To include the inelastic case, we introduce another degree of freedom (for instance, spin) and two channels as $\ket{\alpha}$ and $\ket{\beta}$. Then any state can be written as:
    \begin{equation}
        \ket{\Phi}=\ket{\Phi_{\alpha}}\ket{\alpha}+\ket{\Phi_{\beta}}\ket{\beta}
    \end{equation}
To simplify our calculation, we restrict ourselves to two channels, but cases for more channels can be handled similarly. 

To apply Floquet theory, we assume that the time-dependent scattering wave function is $\ket{\Psi(t)}=e^{-i\epsilon_F t}\ket{\Phi(t)}$, where $\ket{\Phi(t)}$ is a periodic function. The initial conditions can be satisfied by setting $\epsilon_F=\frac{k^2}{2\mu}+E_{\alpha}$ and $\ket{\Phi(t)}\rightarrow \ket{\vec{k}}\ket{\alpha}$ as $t\rightarrow \infty$. Here we've used the assumption that $\braket{\alpha|\hat{V}^{as}(t)|\alpha}=0$ (if not, we can always apply an interaction picture to satisfy this condition). 

Before writing down the Floquet coupled-channel equation, we will explain several notations to be used.
\begin{eqnarray}
\ket{\Phi^n}&=\frac{1}{T}\int_0^T dt e^{i n\omega t}\ket{\Phi(t)}\\
\ket{\Phi^n}&=\ket{\Phi^n_{\alpha}}\ket{\alpha}+\ket{\Phi^n_{\beta}}\ket{\beta}\\
\hat{V}_n&=\frac{1}{T}\int_0^T dt e^{-in \omega t}\hat{V}(t)\\
P_{\alpha}=\ket{\alpha}\bra{\alpha}&,P_{\beta}=\ket{\beta}\bra{\beta}\\
H_{\alpha\beta}=P_{\alpha}H_0P_{\beta}&,H_{\alpha\alpha}=P_{\alpha}H_0P_{\alpha}
\end{eqnarray}
With these preparations, we can finally give the Floquet coupled-channel equation.
\begin{equation}
    \left\{
    \begin{aligned}
    (n\omega+\epsilon_F)\ket{\Phi^n_{\alpha}}\ket{\alpha}=H_{\alpha\alpha}\ket{\Phi^n_{\alpha}}\ket{\alpha}+H_{\alpha\beta}\ket{\Phi^n_{\beta}}\ket{\beta}+\\
    \sum_m V^i_{n-m}\ket{\Phi^m}+V^{as}_{n-m}\ket{\Phi^m_{\alpha}}\ket{\alpha}\\
    (n\omega+\epsilon_F)\ket{\Phi^n_{\beta}}\ket{\beta}=H_{\beta\beta}\ket{\Phi^n_{\beta}}\ket{\beta}+H_{\beta\alpha}\ket{\Phi^n_{\alpha}}\ket{\alpha}+\\
    \sum_m V^i_{n-m}\ket{\Phi^m}+V^{as}_{n-m}\ket{\Phi^m_{\beta}}\ket{\beta}
    \end{aligned}
    \right.
\end{equation}

Comparing these equations with the normal coupled-channel equation in the case of time-independent inelastic scattering, we find that the crucial difference is that the asymptotic term $V^{as}$, which contains the Zeeman term, doesn't vanish at infinity. Therefore we can't solve this equation by directly applying Green's function. 

To solve this equation, we need to diagonalize the asymptotic term first. To simplify the calculation, we assume that the asymptotic term has the Zeeman form, that is, two different constants, $V^{\alpha}_{n}$ and $V^{\beta}_{n}$ for the two channels. Define a matrix $H_e$, which subjects to:
\begin{equation}
    (H_e)^{\alpha,\beta}_{nm}=(n\omega+\epsilon_F)\delta_{nm}-V^{\alpha,\beta}_{n-m}
\end{equation}
Clearly $H_e^{\pm}$ are Hermitian, thus there exist unitary matrixs $A^{\pm}$ which can diagonalize $H_e^{\pm}$:
\begin{equation}
    A^{\pm}H_e^{\pm}(A^{\pm})^{-1}=H_d^{\pm},\quad (H_d^{\pm})_{nm}=E_n^{\pm}\delta_{nm}
\end{equation}
The eigenvectors of $H_e^{\pm}$ are:
\begin{equation}
    \ket{\phi_{\pm}^n}=\sum_m (A^{\pm})_{nm}\ket{\Phi_{\pm}^m}
\end{equation}
And we have several new matrices:
\begin{eqnarray}
    U_{nm}&=\sum_p(A^+)_{np}[(A^-)^{-1}]_{pm}\\
    (V^i_{\alpha\beta})_{nm}&=(A^+)_{np} (V_i)_{p-q} [(A^-)^{-1}]_{qm}\\
    (V^i_{\alpha\alpha})_{nm}&=(A^+)_{np} (V_i)_{p-q} [(A^+)^{-1}]_{qm}\\
    (V^i_{\beta\beta})_{nm}&=(A^-)_{np} (V_i)_{p-q} [(A^-)^{-1}]_{qm}\\
    (V^i_{\beta\alpha})_{nm}&=(A^-)_{np} (V_i)_{p-q} [(A^+)^{-1}]_{qm}
\end{eqnarray}

With these preparation, we can rewrite the coupled-channel equations and eliminate the asymptotic terms.
\begin{equation}
    \left\{
    \begin{aligned}
        E^n_{\alpha}\ket{\phi^n_{\alpha}}\ket{\alpha}=H_{\alpha\alpha}\ket{\phi^n_{\alpha}}\ket{\alpha}+\sum_m U_{nm}H_{\alpha\beta}\ket{\phi_{\beta}^m}\ket{\beta}+\\
        \sum_m(V^i_{\alpha\alpha})_{nm}\ket{\phi_{\alpha}^m}\ket{\alpha}+\sum_m(V^i_{\alpha\beta})_{nm}\ket{\phi_{\beta}^m}\ket{\beta}\\
        E^n_{\beta}\ket{\phi^n_{\beta}}\ket{\beta}=H_{\beta\beta}\ket{\phi^n_{\beta}}\ket{\beta}+\sum_m U^{+}_{nm}H_{\beta\alpha}\ket{\phi_{\alpha}^m}\ket{\alpha}+\\
        \sum_m(V^i_{\beta\beta})_{nm}\ket{\phi_{\beta}^m}\ket{\beta}+\sum_m(V^i_{\beta\alpha})_{nm}\ket{\phi_{\alpha}^m}\ket{\alpha}
    \end{aligned}
    \right.
\end{equation}

Now with the absence of the asymptotic terms, we can obtain the Floquet scattering wave function by applying the Green function. To simplify the expression, we only consider the case where the time-dependent Hamiltonian consists only of Zeeman terms, guaranteeing the elimination of $V^i$ terms. We also assume that $H_{\alpha\alpha}=T+V_s^{\alpha}$ and $H_{\beta\beta}=T+V_s^{\beta}$, where $T$ is kinetic term and $V_s$ is potential energy.

With these assumptions, we can give a simple expression of the Floquet Lippmann-Schwinger equation.

\begin{equation}
    \left\{
    \begin{aligned}
        \ket{\phi^n_{\alpha}}\ket{\alpha}&=A_{n0}^{\alpha}\ket{\vec{k}}\ket{\alpha}+G_0(E_n^{\alpha})\left(V_s^{\alpha}\ket{\phi_{\alpha}^{n}}\ket{\alpha}+\sum_m U_{nm}H_{\alpha\beta}\ket{\phi_{\beta}^m}\ket{\beta}\right)\\
        \ket{\phi^n_{\beta}}\ket{\beta}&=G_0(E_n^{\beta})\left(V_s^{\beta}\ket{\phi_{\beta}^{n}}\ket{\beta}+\sum_m U^{+}_{nm}H_{\beta\alpha}\ket{\phi_{\alpha}^m}\ket{\alpha}\right)
    \end{aligned}
    \right.
\end{equation}

where
\begin{equation}
    G_0(E)=\lim\limits_{\epsilon\rightarrow 0}(E-T\pm\epsilon)^{-1}
\end{equation}

To get the scattering cross section, we have to return to the original time-dependent wave function $\ket{\Psi(t)}$ and calculate the time-average scattering cross section, which is valid when the period of the external field is much shorter than the typical scattering time.

\begin{equation}
    \begin{aligned}
    &\ket{\Psi(t)}\\
    &\sim e^{-i \epsilon_F t}\{e^{ikz}\ket{\alpha}+\\
    &\sum_p e^{-i p\omega t}\sum_n f_{n,p}^{\alpha}\frac{e^{i k_{n\alpha} r}}{r}\ket{\alpha}+
    \sum_p e^{-i p\omega t}\sum_n f_{n,p}^{\beta}\frac{e^{i k_{n\beta}r}}{r}\ket{\beta}
    \}
    \end{aligned}
\end{equation}
where
\begin{eqnarray}
    f_{n,p}^{\alpha}&=(A^{\alpha})^{-1}_{pn}(\braket{\vec{k_{n\alpha}}|V_s^{\alpha}|\phi_{\alpha}^n}+\sum_m U_{nm}\braket{\vec{k_{n\alpha}}|H_{\alpha\beta}|\phi_{\beta}^m})\\
    f_{n,p}^{\beta}&=(A^{\beta})^{-1}_{pn}(\braket{\vec{k_{n\beta}}|V_s^{\beta}|\phi_{\beta}^n}+\sum_m U^+_{nm}\braket{\vec{k_{n\beta}}|H_{\beta\alpha}|\phi_{\alpha}^m})
\end{eqnarray}

Before any further deduction, we must specify the explicit form of $A^{\alpha,\beta}$. This can be achieved by resorting to another interaction picture transformation. Let $W=exp(i\int_0^{t} V^{as}(\tau) d\tau)$, and $\ket{\phi(t)}=W\ket{\Phi(t)}$, then we have:
\begin{equation}
    i\frac{d}{dt}\ket{\phi(t)}=W H^0 W^{-1}\ket{\phi(t)}
\end{equation}
Transforming this equation into Floquet form, we can naturally get rid of the asymptotic terms. Therefore the Floquet component of $\ket{\phi(t)}$ are exactly $\ket{\phi_{\alpha}^n}$ and $\ket{\phi_{\beta}^n}$ as we have defined previously.
Thus we have an explicit form of $A^{\alpha,\beta}$:
\begin{equation}
    (A^{\alpha,\beta})_{nm}=W^{n-m}=\frac{1}{T}\int_0^{T}exp(i\int_0^{t} V_{\alpha\alpha,\beta\beta}^{as}(\tau) d\tau)e^{-i (n-m) \omega t} dt
\end{equation}
If we apply a interaction picture to eliminate the asymptotic term in $\alpha$ channel (This can be done because we assume time-dependent terms in the Hamiltonian are Zeeman-like terms) and assume that the asymptotic terms only have one mode, i.e. $V^{as}(t)=V_0 cos(\omega t)$, then we have:
\begin{equation}
    A^{\alpha}_{nm}=\delta_{nm},\: A^{\beta}_{nm}=J_{n-m}\left(\frac{V_0}{\omega}\right)
\end{equation}
where $J$ is the Bessel function.

Substitute these results back into Eqn. (29), we get:
\begin{equation}
    \begin{aligned}
        &\ket{\Psi(t)}\\
        &\sim  e^{-i \epsilon_F t}\{e^{ikz}\ket{\alpha}+\\
        &\sum_n e^{-i n\omega t} g_{n}^{\alpha}\frac{e^{i k_{n\alpha} r}}{r}\ket{\alpha}+\\
        &\sum_n e^{-i n\omega t}\exp(-i\frac{V_0}{\omega}\sin(\omega t)) g_{n}^{\beta}\frac{e^{i k_{n\beta}r}}{r}\ket{\beta}
        \}
    \end{aligned}
\end{equation}
where
\begin{eqnarray}
    g_{n}^{\alpha}=&\braket{\vec{k_{n\alpha}}|V_s^{\alpha}|\phi_{\alpha}^n}+\sum_m (A^{\beta})^{-1}_{nm}\braket{\vec{k_{n\alpha}}|H_{\alpha\beta}|\phi_{\beta}^m}\\
    g_{n}^{\beta}=&\braket{\vec{k_{n\beta}}|V_s^{\beta}|\phi_{\beta}^n}+\sum_m (A^{\beta})_{nm}\braket{\vec{k_{n\beta}}|H_{\beta\alpha}|\phi_{\alpha}^m}
\end{eqnarray}

Therefore the inelastic scattering cross section can be expressed in the following form (detailed derivation can be found in Appendix.A):
\begin{equation}
    \begin{aligned}
    \sigma_{\alpha\rightarrow\beta}&=\sum_n \sigma_{\alpha\rightarrow\beta}^n\\
    \sigma_{\alpha\rightarrow\beta}^n&=\int Re\left\{\sum_n |g_{n}^{\beta}|^2 \frac{k_{n,\beta}}{k}\right\} d\Omega
    \end{aligned}
\end{equation}

\begin{equation}
    \begin{aligned}
    \sigma_{\alpha\rightarrow\alpha}&=\sum_n \sigma_{\alpha\rightarrow\alpha}^n\\
    \sigma_{\alpha\rightarrow\alpha}^n&=\int Re\left\{\sum_n |g_{n}^{\alpha}|^2 \frac{k_{n,\alpha}}{k}\right\} d\Omega
    \end{aligned}
\end{equation}

\subsection{\label{sec:level2}Scattering resonance and comparison between elastic and inelastic Floquet scattering}

In [1], Smith had developed a formal theory of Floquet Feshbach resonance. It would be natural to assume that other Floquet scattering may follow the same scheme. However, as we will demonstrate in this section, there is a crucial difference between elastic and inelastic Floquet scattering process in terms of scattering resonance. (Here elastic process refers to those through which the internal degree of freedom remains the same, but the energy of the particle may change.)

To illustrate this difference, we first apply a interaction picture to eliminate the asymptotic term in $\alpha$ channel . Meanwhile, to include the scattering resonance effect, we introduce another channel $\ket{\gamma}$ which only supports bound states. Then the coupled-channel equation can be written as:
\begin{equation}
    \left\{
    \begin{aligned}
    (n\omega+\epsilon_F)\ket{\Phi^n_{\alpha}}\ket{\alpha}=H_{\alpha\alpha}\ket{\Phi^n_{\alpha}}\ket{\alpha}+H_{\alpha\beta}\ket{\Phi^n_{\beta}}\ket{\beta}+\\
    \sum_m (V^i_{\alpha\alpha})_{n-m}\ket{\Phi^m}\ket{\alpha}+(V^i_{\alpha\beta})_{n-m}\ket{\Phi^m}\ket{\beta}\\
    (n\omega+\epsilon_F)\ket{\Phi^n_{\beta}}\ket{\beta}=H_{\beta\beta}\ket{\Phi^n_{\beta}}\ket{\beta}+H_{\beta\alpha}\ket{\Phi^n_{\alpha}}\ket{\alpha}+\\
    \sum_m (V^i_{\beta\alpha})_{n-m}\ket{\Phi^m}\ket{\alpha}+(V^i_{\beta\beta})_{n-m}\ket{\Phi^m}\ket{\beta}+V^{as}_{n-m}\ket{\Phi^m_{\beta}}\ket{\beta}
    \end{aligned}
    \right.
\end{equation}
where $V^i$ mainly consists of the effective Hamiltonian due to interaction with $\ket{\gamma}$ channel. Suppose the equation for $\ket{\gamma}$ channel has the following form:
\begin{equation}
\begin{aligned}
    (n\omega+\epsilon_F)\ket{\Phi_{\gamma}^n}\ket{\gamma}&=H_{\gamma\gamma}\ket{\Phi_{\gamma}^n}\ket{\gamma}+\sum_m V^{\gamma}_{nm}\ket{\Phi_{\gamma}^m}\ket{\gamma}+K\ket{\Phi^n}\\
    where\quad\ket{\Phi^n}&=\ket{\Phi^n_{\alpha}}\ket{\alpha}+\ket{\Phi^n_{\beta}}\ket{\beta}
\end{aligned}
\end{equation}
then we have:
\begin{equation}
    V^i=K^{+}G_{\gamma}K
\end{equation}
where $G_{\gamma}$ satisfies:
\begin{equation}
    \sum_m (G_{\gamma})_{nm}[(m\omega+\epsilon_F-H_{\gamma\gamma})\delta_{mp}-V_{mp}^{\gamma}]=\delta_{np}
\end{equation}
namely, $G_{\gamma}$ is the Green's function.
\subsubsection{Elastic case}
We first consider the case of elastic scattering, in which we can get rid of $\ket{\beta}$ channel.

In this case, Eqn.(40) can be written as:
\begin{equation}
\begin{aligned}
    (n\omega+\epsilon_F)\ket{\Phi^n_{\alpha}}\ket{\alpha}=H_{\alpha\alpha}\ket{\Phi^n_{\alpha}}\ket{\alpha}+\sum_m K^+(G_{\gamma})_{n-m}K\ket{\Phi_{\alpha}^m}\ket{\alpha}
\end{aligned}
\end{equation}

Then we get a expression of the Floquet Lippmann-Schwinger equation:
\begin{equation}
\begin{aligned}
    \ket{\Phi_{\alpha}^n}\ket{\alpha}=&\delta_{n0}\ket{\vec{k}}\ket{\alpha}+\\
    &G_0(n\omega+\epsilon_F)(V_s^{\alpha}\ket{\Phi_{\alpha}^n}\ket{\alpha}+\sum_m K^+(G_{\gamma})_{nm}K\ket{\Phi_{\alpha}^m}\ket{\alpha})
\end{aligned}
\end{equation}
and the corresponding scattering cross sections are:
\begin{eqnarray}
    g_n^{\alpha}&=\braket{\vec{k_{n\alpha}}|V_s^{\alpha}|\Phi_{\alpha}^n}+\sum_m \braket{\vec{k_{n\alpha}}|K^+(G_{\gamma})_{nm}K|\Phi_{\alpha}^m}\\
    \sigma_{\alpha\rightarrow\alpha}&=\sum_n \sigma_{\alpha\rightarrow\alpha}^n\\
    \sigma_{\alpha\rightarrow\alpha}^n&=\int \sum_n |g_{n}^{\alpha}|^2 \frac{k_{n,\alpha}}{k} d\Omega
\end{eqnarray}

Before moving on, we make a further assumption that $V^{\gamma}$ is small compared to $\omega$. This is, of course, not always the case, but we can best illustrate the difference between elastic and inelastic Floquet scattering in this case.

We expand $G_{\gamma}$ in series of $V^{\gamma}$:
\begin{equation}
\begin{aligned}
    (G_{\gamma})_{nm}=&\frac{1}{n\omega+\epsilon_F-H_{\gamma\gamma}}\delta_{nm}+\\
    &\frac{1}{n\omega+\epsilon_F-H_{\gamma\gamma}}V^{\gamma}_{nm}\frac{1}{m\omega+\epsilon_F-H_{\gamma\gamma}}+\\
    &\sum_p\frac{1}{n\omega+\epsilon_F-H_{\gamma\gamma}}V^{\gamma}_{np}\frac{1}{p\omega+\epsilon_F-H_{\gamma\gamma}}V^{\gamma}_{pm}\frac{1}{m\omega+\epsilon_F-H_{\gamma\gamma}}+\\
    &......
\end{aligned}
\end{equation}

If we consider the simplest case where the external field has only one mode, namely, $V^{\gamma}_{nm}$ is non-zero only when $|n-m|\leq 1$, then the leading term of $G_{n,(n+N)}$ is:
\begin{equation}
\begin{aligned}
    \widetilde{(G_{\gamma})}_{n,n+N}=\frac{1}{n\omega+\epsilon_F-H_{\gamma\gamma}}V^{\gamma}_{+1}\frac{1}{(n+1)\omega+\epsilon_F-H_{\gamma\gamma}}V^{\gamma}_{+1}\\
    ...V^{\gamma}_{+1}\frac{1}{(n+N)\omega+\epsilon_F-H_{\gamma\gamma}}
\end{aligned}
\end{equation}
where $V_{+1}^{\gamma}=V_{n,n+1}^{\gamma}=...=V_{N-1,N}^{\gamma}$

Therefore, the leading term of $g_n^{\alpha}$ is:
\begin{equation}
    \widetilde{g_n^{\alpha}}=\braket{\vec{k_{n\alpha}}|V_s^{\alpha}|\Phi_{\alpha}^n}+\braket{\vec{k_{n\alpha}}|K^+\widetilde{(G_{\gamma})}_{n0}K|\vec{k}}
\end{equation}

Physically, this result means that during the scattering process, the particle can only gain energy from the external field step by step. For example, if the particle is to reach a final state with energy $\epsilon+10\omega$, it must go through all the intermediate states with energy $\epsilon+\omega$, $\epsilon+2\omega$, etc. Therefore, if $\epsilon+\omega$ approaches the energy of a bound state of $H_{\gamma\gamma}$, all the scattering cross section $\sigma_{\alpha\rightarrow\alpha}^{n}$ would resonant simultaneously.

\subsubsection{Inelastic case}
To simplify the equation, we only consider the process in which particles can only transit from channel $\alpha$ to $\beta$ through channel $\gamma$. In this case, the coupled-channel equation can be written as:
\begin{equation}
    \left\{
    \begin{aligned}
    &(n\omega+\epsilon_F)\ket{\Phi^n_{\alpha}}\ket{\alpha}=H_{\alpha\alpha}\ket{\Phi^n_{\alpha}}\ket{\alpha}+\\
    &\sum_m (V^i_{\alpha\alpha})_{n-m}\ket{\Phi^m_{\alpha}}\ket{\alpha}+(V^i_{\alpha\beta})_{n-m}\ket{\Phi^m_{\beta}}\ket{\beta}\\
    &(n\omega+\epsilon_F)\ket{\Phi^n_{\beta}}\ket{\beta}=H_{\beta\beta}\ket{\Phi^n_{\beta}}\ket{\beta}\\
    &\sum_m (V^i_{\beta\alpha})_{n-m}\ket{\Phi^m_{\alpha}}\ket{\alpha}+(V^i_{\beta\beta})_{n-m}\ket{\Phi^m_{\beta}}\ket{\beta}+V^{as}_{n-m}\ket{\Phi^m_{\beta}}\ket{\beta}
    \end{aligned}
    \right.
\end{equation}
where $V^i=K^+G_{\gamma}K$.

Similar to the elastic case, the leading term of $g_n^{\beta}$ is:
\begin{equation}
    \widetilde{g_n^{\beta}}=\sum_p A^{\beta}_{np}\braket{\vec{k_{n\beta}}|K^+_{\beta}\widetilde{(G_{\gamma})}_{p0}K_{\alpha}|\vec{k}}
\end{equation}
where $K_{\beta}^+=P_{\beta}K^+$ and $K_{\alpha}=K P_{\alpha}$.

Here we also consider the simplest case where the external field has only one mode.

We can physically interpret this result as, in inelastic scattering, the particle can reach the final state by skipping many intermediate states. This results embedded in the nature of the matrix $A^{\beta}$ which can connect Floquet blocks with distance larger than 1 even the external field has only one mode. 

This property lead to the consequence that the scattering cross section $\sigma_{\alpha\rightarrow\beta}^{n}$ may not resonant simultaneously, which is in contrast with the results in elastic case.

\section{\label{sec:level1}Numerical implementation}
In this section, we will present our detailed numerical implementation for the inelastic scattering cross section in the main text. We first calculate the matrix elements of the Floquet Hamiltonian and diagonalize it in A. Then we will derive the coupled channel equation in B. And in C, by matching the wave function with the boundary condition, we give a explicit expression for the scattering cross section. Finally in D, we discuss the influence of thermal distribution of the collision energy $E_{\rm{in}}$.

\subsection{\label{sec:level2}Diagonalization of the Floquet Hamiltonian}
For nonreactive collisions of diatomic molecules in the electronic states of $^{2} \Sigma$ and  $^{3} \Sigma$ symmetry with structureless atoms in temporally varying magnetic field, the total Hamiltonian (in units of $\hbar$)
\begin{equation}
\hat{H}=-\frac{1}{2\mu R}\frac{\partial^2}{\partial R^2}R+\frac{\hat{l} ^2}{2\mu R^2}+\hat{V}(\boldsymbol{R},\boldsymbol{r})+\hat{H}_{\rm{as}},
\end{equation}
where $\boldsymbol{R}$ is the atom-molecule separation vector, $\mu$ is the reduced mass of the collision complex, $\hat{l}$ is the orbital angular momentum for the collision, and $\boldsymbol{r}$ is the separation vector between the atoms in the diatomic molecule. For diatomic molecules in its vibrational ground state, the distance between the atoms can be viewed as a constant, so the interaction potential $\hat{V}(\boldsymbol{R},\boldsymbol{r})$ can be written as $\hat{V}(R,\theta)$, where $\theta$ is the angle between $\boldsymbol{R}$ and $\boldsymbol{r}$.

The asymptotic  Hamiltonian $\hat{H}_{\rm{as}}$ describes the rotational structure of a $^{2} \Sigma$ molecule in its electronic and vibrational ground state in the presence of a time-varying field $B(t)$ with period $T=2\pi/\omega_B$~\cite{spinde,diatom},
\begin{equation}
\hat{H}_{\rm{as}}=B_e \hat{N} ^2+\gamma \hat{N}\cdot \hat{S}+2\mu_0 \hat{B}(t)\cdot\hat{S}, 
\end{equation}
where $B_e$ is the rotational constant $\hat{N}$ is the rotational angular momentum, $\hat{S}$ is the electron spin, $\mu_0$ is the Bohr magnetom and $\gamma$ is a phenomenological spin-rotation constant. For a $^3\Sigma$ molecule, the asymptotic Hamiltonian can be obtained by adding to Eq.(2) a term describing the spin-spin interaction~\cite{diatom}
\begin{equation}
\hat{V}_{SS}=\frac{2}{3}\lambda_{SS}\left(\frac{24\pi}{5}\right)^{1/2} \sum_{q=-2}^{2} (-)^q Y_{2,-q}(\hat{r})[\hat{S} \otimes \hat{S}]_{q}^{(2)},
\end{equation}
where $\lambda_{SS}$ is the spin-spin constant and $[\hat{S} \otimes \hat{S}]_{q}^{(2)}$ is a spherical tensor product.

By expanding the Hilbert space to include time $t$ as a operator, we get the so-called total Floquet Hamiltonian
\begin{equation}
\hat{H}_{\rm{F}}=-\frac{1}{2\mu R}\frac{\partial^2}{\partial R^2}R+\frac{\hat{l} ^2}{2\mu R^2}+\hat{V}(\boldsymbol{R},\boldsymbol{r})+\hat{H}_{\rm{as}}-i\frac{\partial}{\partial t}.
\end{equation}
Define $H_{\rm{asF}}=H_{\rm{as}}-i\frac{\partial}{\partial t}$. Recalling from Floquet theory, we have $\langle \alpha n|H_{\rm{asF}}|\beta m\rangle=H^{n-m}_{\alpha\beta}+n\omega_B\delta_{\alpha\beta}\delta_{nm}$, where $\alpha,\beta$ denote different channels and $H^{n-m}_{\alpha\beta}=\frac{1}{T}\int_{0}^{T}dt \braket{\alpha|H_{\rm{as}}|\beta}e^{-i(n-m)\omega_B t}$. Therefore we can calculate the matrix elements of $H_{\rm{asF}}$ under the basis $\ket{NSJM_Jn}$ ($\hat{J}=\hat{N}+\hat{S}$, and $\langle t|n\rangle=e^{in\omega_B t}$ denotes the Floquet index).

We present the explicit expression for the matrix elements here:
\begin{eqnarray}
    &&\braket{NSJM_Jn|\gamma \hat{N}\cdot \hat{S}|N'SJ'M_J'm}\nonumber\\
    &&=\delta_{N,N'}\delta_{J,J'}\delta_{M_J,M_J'}\delta_{n,m}(-1)^{N+J+S}[N(N+1)(2N+1)S(S+1)(2S+1)]^{1/2}\left\{
    \begin{matrix}
S& N& J \\
N& S& 1 \\
\end{matrix}
    \right\},
\end{eqnarray}
\begin{eqnarray}
    &&\braket{NSJM_Jn|\hat{V}_{SS}|N'SJ'M_J'm}\nonumber\\
    &&=\delta_{J,J'}\delta_{M_J,M_J'}\delta_{n,m}\frac{2\sqrt{30}}{3}\lambda_{SS}(-1)^{J+N'+N+S}[(2N+1)(2N'+1)]^{1/2}
    \left(
    \begin{matrix}
    N& 2& N'\\
    0& 0& 0 \\
    \end{matrix}
    \right)
    \left\{
    \begin{matrix}
    S& N'& J\\
    N& S & 2 \\
    \end{matrix}
    \right\},
    \nonumber\\
    &&
\end{eqnarray}
\begin{eqnarray}
    &&\braket{NSJM_Jn|2\mu_0 B(t)\cdot\hat{S}|N'SJ'M_J'm}\nonumber\\
    &=&\delta_{N,N'}\delta_{M_J,M_J'}2\mu_0B_{n-m}(-1)^{N+S-M_J+1}[S(S+1)(2S+1)(2J+1)(2J'+1)]^{1/2}
     \left(
    \begin{matrix}
    J& 1& J'\\
    0& 0& 0 \\
    \end{matrix}
    \right)
    \left\{
    \begin{matrix}
    S& J'& N\\
    J& S & 1 \\
    \end{matrix}
    \right\},
    \nonumber\\
    &&
\end{eqnarray}
where $B_{n-m}=\frac{1}{T}\int_{0}^{T}dtB(t)e^{-i(n-m)\omega_B t}$.


By diagonalizing the matrix $\langle \alpha n|H_{\rm{asF}}|\beta m\rangle$, we can get the eigenfunction $\Phi_{\alpha K}$ of $H_{\rm{asF}}$ with corresponding eigenvalue $\epsilon_{\alpha K}$,
\begin{equation}
    \Phi_{\alpha K}=\sum_{N,S,J,M_J,n} W_{\alpha K,NSJM_Jn} |NSJM_Jn\rangle.
\end{equation}

\subsection{\label{sec:level2}Coupled-channel equation}
If we expand the total scattering wave function as:
\begin{equation}
\Psi =R^{-1}\sum_{\alpha Klm_l}F^{lm_l}_{\alpha K}(R)\Phi_{\alpha K}|{l,m_l}\rangle,
\end{equation}
then we can derive the so-called coupled-channel equation for the coefficients $F^{lm_l}_{\alpha K}(R)$ from $\hat{H}_F\Psi=E_F\Psi$,
\begin{equation}
    \frac{d^2}{dR^2}F^{l'm_l'}_{\alpha' K'}+\sum_{\alpha K l m_l}[(k^2_{\alpha K}-\frac{l(l+1)}{R^2})\delta_{\alpha',\alpha}\delta_{K',K}\delta_{l',l}\delta_{m_l',m_l}-2\mu V_{\alpha' K' l' m_{l}',\alpha K l m_{l}}]F^{l m_l}_{\alpha K}=0,
\end{equation}
where
\begin{eqnarray}
k^{2}_{\alpha K}=2\mu(E_{F}-\epsilon_{\alpha K}),
\end{eqnarray}
\begin{eqnarray}
V_{\alpha' K' l' m_{l}',\alpha K l m_{l}}(R)=\langle l'm_l'|\langle \Phi_{\alpha' K'}|V(R,\theta)|\Phi_{\alpha K}\rangle|lm_l\rangle,
\end{eqnarray}
$E_{F}$ is the total Floquet energy and $\epsilon_{\alpha K}$ is the asymptotic energy of $\Phi_{\alpha K}$.

\subsection{\label{sec:level2}Scattering cross section}
For the scattering wave function $\Psi =R^{-1}\sum_{\alpha' K' l' m_l'}F^{l' m_l'}_{\alpha' K'}(R)\Phi_{\alpha' K'}|{l',m_l'}\rangle$ from the initial state $\Phi_{\alpha K}|{l,m_{l}}\rangle$, the radial expansion coefficients are subject to the boundary conditions
\begin{equation}
F^{l' m_{l}'}_{\alpha' K'}(R\to\infty)\sim \frac{sin(k_{\alpha K}R-l\pi/2)}{\sqrt{k_{\alpha K}}}\delta_{\alpha, \alpha'}\delta_{K,K'}\delta_{l,l'}\delta_{m_l, m_{l}'}-\frac{e^{i(k_{\alpha' K'}R-l\pi/2)}}{2i\sqrt{k_{\alpha' K'}}}T_{\alpha' K' l' m_{l}',\alpha K l m_{l}},
\end{equation}
gives the ${T}$-matrix for transitions between eigenstates labeled by indexes ($l,m_{l},\alpha,K$). 

Define an matrix $\bm{F}$ in which the $i^{th}$ column represents the radial coefficients of scattering wavefunction from the $i^{th}$ initial state ($l_i,m_{l_i},\alpha_i,K_i$). In other word, the matrix element in the $i^{th}$ column are the solution of Eq.(63) propagating from the $i^{th}$ initial state. Then the boundary condition can be written as:
\begin{equation}
    \bm{F}(R\rightarrow \infty)=\bm{J}(R)-\frac{1}{2}\bm{H^+}(R)\bm{T},
\end{equation}
where $\bm{J}$ and $\bm{H}$ are diagonal matrix with diagonal elements of the Ricatti-Bessel function and the Ricatti-Hankel function respectively, namely,
\begin{equation}
\begin{aligned}
    (\bm{J}(R))_{\alpha' K' l' m_{l}',\alpha K l m_{l}}&=\delta_{\alpha, \alpha'}\delta_{K,K'}\delta_{l,l'}\delta_{m_l, m_{l}'}k^{1/2}_{\alpha K}Rj_{l}(k_{\alpha K}R),\\
    (\bm{N}(R))_{\alpha' K' l' m_{l}',\alpha K l m_{l}}&=\delta_{\alpha, \alpha'}\delta_{K,K'}\delta_{l,l'}\delta_{m_l, m_{l}'}k^{1/2}_{\alpha K}Rn_{l}(k_{\alpha K}R),\\
    \bm{H}^{\pm}&=\bm{J}\pm i\bm{N},
\end{aligned} 
\end{equation}
and
\begin{equation}
    \bm{T}=4i\mu \int_0^{\infty}\bm{J}^T(R)\bm{V}(R)\bm{F}(R)dR.
\end{equation}

The integral cross sections are evaluated from the ${T}$-matrix elements as
\begin{equation}
\sigma_{\alpha K\to \alpha' K'}=\frac{\pi}{k^2_{\alpha K}}\sum_M \sum_{ll'}\sum_{m_l m_{l'}}\left|T_{\alpha' K' l' m_{l}',\alpha K l m_{l}}\right|^2.
\end{equation}
The total angular momentum projection $M$ is conserved during the collision so that computations can be carried out independently of different values of $M$.

\subsection{\label{sec:level2}The impact of thermal distribution}
The collision energy $E_{\rm{in}}$ of the system approximately satisfies Maxwellian distribution characterized by the kinetic temperature, and the scattering cross section should consider the impact of thermal averaging in a gas. As mentioned above, the ${T}$-matrix is defined by $\bm{T}=4i\mu \int_0^{\infty}\bm{J}^T(R)\bm{V}(R)\bm{F}(R)dR$. By applying the distorted wave Born approximation, the elements of ${T}$-matrix are
\begin{equation}
    T_{\alpha' K' l' m_{l}',\alpha K l m_{l}}=4i\mu \int_0^{\infty}j_{l'}(k_{\alpha' K'}R)V_{\alpha' K' l' m_{l}',\alpha K l m_{l}}(R)j_{l}(k_{\alpha K}R)dR,
\end{equation}
where $k^2_{\alpha K}=2\mu E_{\rm{in}}$ and $k^2_{\alpha' K'}=2\mu E_{\rm{out}}$ are the wave vector before and after the collision, respectively. For ultracold collisions, $k_{\alpha K}R\ll1$ within the effective range of interaction potential $\bm{V}(R)$, thus we have $j_{l}(k_{\alpha K}R)\to \frac{1}{\sqrt{k_{\alpha K}}(2l+1)!!}(k_{\alpha K}R)^{l+1}$. Furthermore, because the Zeeman separation is much larger than the collision energy $E_{\rm{in}}$ considered here, $E_{\rm{out}}$ and $k_{\alpha' K'}$ can be regarded as irrelevant to $E_{\rm{in}}$ and only affected by magnetic field.

\begin{figure*}[t]
    \centering
    \includegraphics[width=12.0cm]{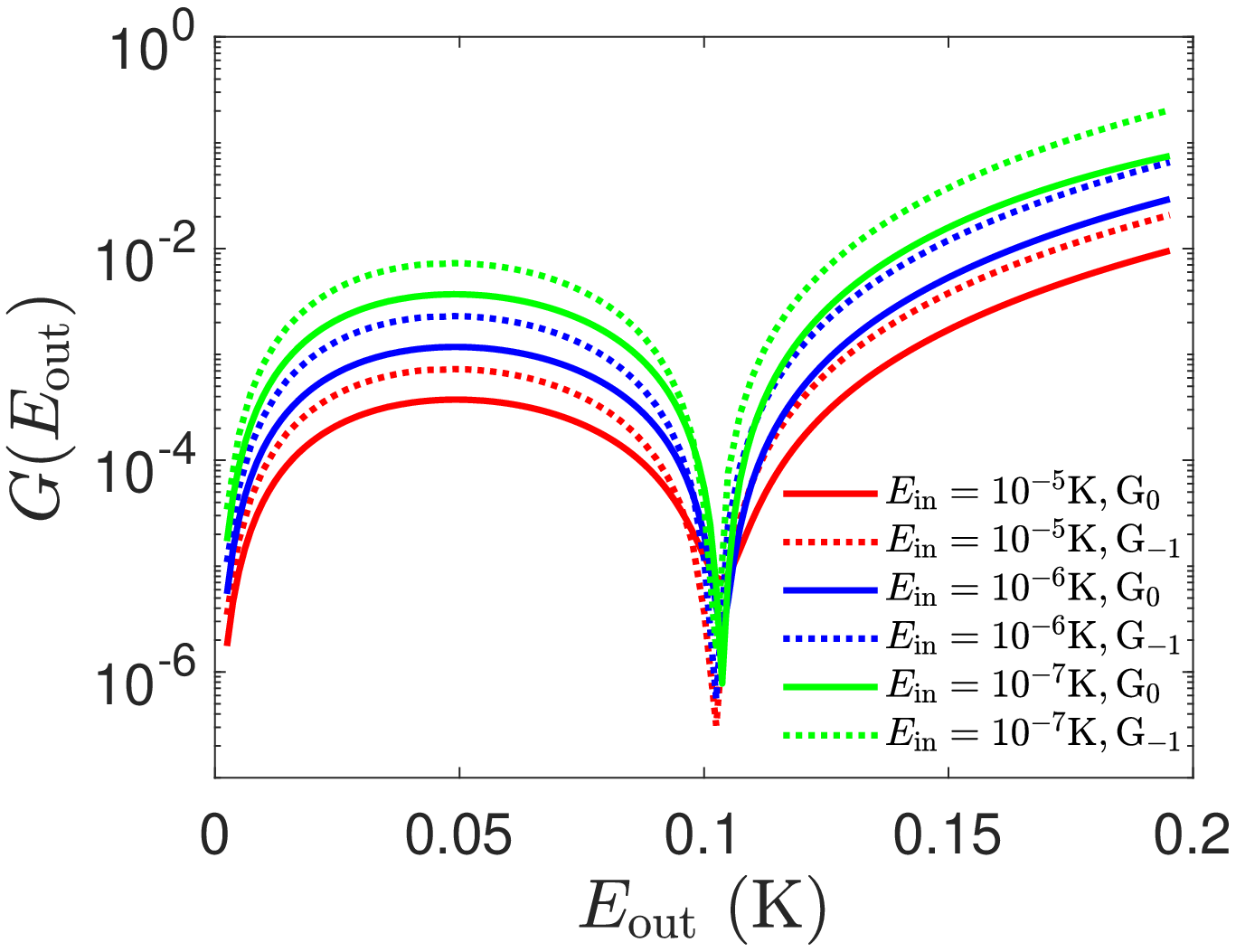}
    \caption{ \label{fig:flc}The transition probabilities $G_0(E_{\rm{out}})$ and $G_{-1}(E_{\rm{out}})$ as a function of the exit kinetic energy $E_{\rm{out}}$ with the collision energy $E_{\rm{out}}=10^{-5}~\rm{K}$~(red), $E_{\rm{out}}=10^{-6}~\rm{K}$~(blue) and $E_{\rm{out}}=10^{-7}~\rm{K}$~(green).}
\end{figure*}
Therefore Eq. (71) can be written as 
\begin{equation}
    T_{\alpha' K' l' m_{l}',\alpha K l m_{l}}=\frac{4i\mu}{(2l+1)!!}k^{l+1/2}_{\alpha K} \int_0^{\infty}j_{l'}(k_{\alpha' K'}R)V_{\alpha' K' l' m_{l}',\alpha K l m_{l}}(R)R^{l+1}dR.
\end{equation}
Notice that the distribution of $E_{\rm{in}}$ only influence the scalar factor $k^{l+1/2}_{\alpha K}$. Since the relative values of transition probabilities $G_0(E_{\rm{out}}),G_{-1}(E_{\rm{out}})$ are equal to cross sections $\sigma_0, \sigma_{-1}$ when only the static field is applied, according to Eq. (70), we have $G_0(E_{\rm{out}}),G_{-1}(E_{\rm{out}})\propto (E_{\rm{in}})^{l-1/2}$ while their variation with the exit kinetic energy $E_{\rm{out}}$ is independent of $E_{\rm{in}}$. In particular, for the inelastic collision $\rm{^3He}~$$\rm{+\,^{17}O_2}$$(M_J=+1)$ $\to\,\rm{^3He}~$$\rm{+\,^{17}O_2}$$(M_J=0,-1)$ considered in the main text, the leading contribution of cross section is s-wave scattering which will undergo $l=0\to l'=2$, hence the transition probabilities are proportional to $(E_{\rm{in}})^{-1/2}$.

Fig.~1 illustrates the transition probabilities $G_0(E_{\rm{out}})$ and $G_{-1}(E_{\rm{out}})$ as a function of $E_{\rm{out}}$ with different collision energies $E_{\rm{in}}$. Note that the shapes of $G_0(E_{\rm{out}})$ and $G_{-1}(E_{\rm{out}})$ are almost unaffected by the variation of $E_{\rm{in}}$ and the zero points in different situations correspond to the same $E_{\rm{out}}$, thus the selective quantum Zeno effect is still valid when we consider the thermal distribution of $E_{\rm{in}}$.

\section{\label{sec:level1}Calculation OF scattering rates via Kofman-Kurizki (KK) scheme}
As discussed in the main text, in the presence of the magnetic pulse train $B(t)=B_0+\mathcal{B}(t)$ with period $T=2\pi/\omega_B$, the total energy will change from $E_{\rm{in}}+hB_0+h\mathcal{B}(t)$ to $E_{\rm{in}}+hB_0-K\omega_B$ (or $E_{\rm{in}}+hB_0-h\mathcal{B}(t)-L\omega_B$) after the inelastic scattering $|1,0\rangle \to |0,K\rangle$ (or $|1,0\rangle \to |-1,L\rangle$). In this case, we use the asymptotic molecular states $|1,0\rangle$, $|0,K\rangle$ and $|-1,L\rangle$ to label distinct levels whose total energies are $E_{\rm{in}}+hB_0+h\mathcal{B}(t)$, $E_{\rm{in}}+hB_0-K\omega_B$ and $E_{\rm{in}}+hB_0-h\mathcal{B}(t)-L\omega_B$, respectively. Here we omit the constant energy $E_{\rm{in}}$ and the effective Hamiltonian can be written as $\hat{H}(t)=\hat{H}_{0}+\hat{V}+H_{1}(t)$, where
\begin{eqnarray}
\hat{H}_{0}= hB_0|1,0\rangle\langle 1,0|+ \sum_{K}(hB_0-K\omega_{\mathrm{B}})|0,K\rangle\langle 0,K|+\sum_{L}(hB_0-L\omega_{\mathrm{B}})|-1,L\rangle\langle -1,L|,
\end{eqnarray}
\begin{eqnarray}
\hat{V}=\sum_{K}V_{0}(hB_0-K\omega_{\mathrm{B}})|1,0\rangle\langle 0,K|+\sum_{L}V_{-1}(2hB_0-L\omega_{\mathrm{B}})|1,0\rangle\langle -1,L|+\mathrm{H.c.},
\end{eqnarray}
\begin{equation}
\hat{H}_{1}(t)=h\mathcal{B}(t)|1,0\rangle\langle 1,0|-\sum_{L}h\mathcal{B}(t)|-1,L\rangle\langle -1,L|,
\end{equation}
The wave function of the system at time $t$ is
\begin{eqnarray}
\Psi(t)&=&c_1(t)e^{-ihB_0t-i\int_0^t h\mathcal{B}(t')dt'}|1,0\rangle+\nonumber\\
&\sum_K& c_{0,K}(t)e^{-i(hB_0-K\omega_{\mathrm{B}} )t}|0,K\rangle+\sum_L \!c_{\!-\!1,L}(t)e^{-i(hB_0-L\omega_{\mathrm{B}} )t+i\int_0^t h\mathcal{B}(t')dt'}\!|\!-\!1,\!L\rangle\!,
\end{eqnarray}
the initial condition being $\Psi(0)=|1,0\rangle$. Substituting Eq. (76) into the Schrodinger equation $i\frac{\partial}{\partial t}\Psi(t)=\hat{H}(t)\Psi(t)$, we obtain
\begin{eqnarray}
c_{0,K}(t)=-i\int_0^t dt' V_{0}(hB_0-K\omega_{\mathrm{B}})c_1(t')e^{-iK\omega_{\mathrm{B}} t'-i\int_0^{t'} h\mathcal{B}(t'') dt''},
\end{eqnarray}
\begin{eqnarray}
c_{-1,L}(t)=-i\int_0^t dt' V_{-1}(2hB_0-L\omega_{\mathrm{B}})c_1(t')e^{-iL\omega_{\mathrm{B}} t'-i\int_0^{t'} 2h\mathcal{B}(t'') dt''},
\end{eqnarray}
and
\begin{eqnarray}
\dot{c_1}(t)=-\int_0^t dt' [ &\epsilon&_0^{*}(t) \epsilon_0(t')\sum_K |V_{0}(hB_0-K\omega_{\mathrm{B}})|^2e^{iK\omega_{\mathrm{B}}(t-t')}+\nonumber\\
&\epsilon&_{-1}^{*}(t) \epsilon_{-1}(t')\sum_L |V_{-1}(2hB_0-L\omega_{\mathrm{B}})|^2e^{iL\omega_{\mathrm{B}}(t-t')}] c_1(t'),
\end{eqnarray}
with $\epsilon_0(t)=\exp[-i\int_0^t h\mathcal{B}(t')dt']$ and $\epsilon_{-1}(t)=\exp[-i\int_0^t 2h\mathcal{B}(t)dt']$. Calculation results demonstrates that the inelastic cross sections are sufficiently small (about $10^{-3}~\angstrom^2$) within the range of magnetic field considered here, so that the coupling in Eq. (74) is a weak perturbation and $c_1(t)$ varies much slowly with respect to the kernel of Eq. (79). One can thus make the approximation $c_1(t')\approx c_1(t)$ on the right-hand side of Eq. (79). Then we solve Eq. (79) and represent the amplitude modulus of level $|1,0\rangle$ in the form
\begin{eqnarray}
|c_1(t)|=\exp\{-[R_0(t)+R_{-1}(t)]\cdot t/2\},
\end{eqnarray}
where 
\begin{eqnarray}
R_0(t)=2\pi\int_{-\infty}^{+\infty} dE  P_0(E+hB_0) Q_{0}(E,t),
\end{eqnarray}
\begin{eqnarray}
R_{-1}(t)=2\pi\int_{-\infty}^{+\infty} dE  P_{-1}(E+hB_0) Q_{-1}(E,t),
\end{eqnarray}
with $P_0(E)=\sum_K |V_{0}(hB_0-K\omega_{\mathrm{B}})|^2\delta(E-hB_0+K\omega_{\mathrm{B}}),~P_{-1}(E)=\sum_L |V_{-1}(2hB_0-L\omega_{\mathrm{B}})|^2\delta(E-hB_0+L\omega_{\mathrm{B}})$ and $Q_{0}(E,t)=\frac{1}{2\pi t}|\int_0^t \epsilon_0(t')e^{iEt'}dt'|^2,~Q_{-1}(E,t)=\frac{1}{2\pi t}|\int_0^t \epsilon_{-1}(t')e^{iEt'}dt'|^2$. 

Expand $\epsilon_0(t)$ and $\epsilon_{-1}(t)$ as Fourier series, $\epsilon_0(t)=\sum_K \lambda_{0,K} e^{-iK\omega_{\mathrm{B}} t}$ and $\epsilon_{-1}(t)=\sum_L \lambda_{-1,L} e^{-iL \omega_{\mathrm{B}} t}$. For $t \gg \omega^{-1}_{\mathrm{B}}$, Eq. (81) and Eq. (82) are reduced to 
\begin{eqnarray}
R_0(t)=2\pi\sum_K |\lambda_{0,-K}|^2 P_0(hB_0-K\omega_{\mathrm{B}}),
\end{eqnarray}
\begin{eqnarray}
R_{-1}(t)=2\pi\sum_L |\lambda_{-1,-L}|^2 P_{-1}(hB_0-L\omega_{\mathrm{B}}).
\end{eqnarray}
Therefore 
\begin{eqnarray}
|c_1(t)|^2&=&\exp\{-[2\pi\sum_K |\lambda_{0,-K}|^2 P_0(hB_0-K\omega_{\mathrm{B}})+2\pi\sum_L |\lambda_{-1,-L}|^2 P_{-1}(hB_0-L\omega_{\mathrm{B}})]\cdot t\}\nonumber\\
&\approx&1-[2\pi\sum_K |\lambda_{0,-K}|^2 P_0(hB_0-K\omega_{\mathrm{B}})+2\pi\sum_L |\lambda_{-1,-L}|^2 P_{-1}(hB_0-L\omega_{\mathrm{B}})]\cdot t
\end{eqnarray}
and 
\begin{eqnarray}
|c_{0,K}(t)|^2\approx2\pi |\lambda_{0,-K}|^2 t P_0(hB_0-K\omega_{\mathrm{B}}),~|c_{-1,L}(t)|^2\approx2\pi |\lambda_{-1,-L}|^2 t P_{-1}(hB_0-L\omega_{\mathrm{B}}).
\end{eqnarray}
As a result, the scattering rates of $|1,0\rangle \to |0,K\rangle$ and $|1,0\rangle \to |-1,L\rangle$ satisfies $\partial_t|c_{0,K}(t)|^2\propto |\lambda_{0,-K}|^2|V_{0}(hB_0-K\omega_{\mathrm{B}})|^2$ and $\partial_t|c_{-1,L}(t)|^2\propto|\lambda_{-1,-L}|^2|V_{-1}(2hB_0-L\omega_{\mathrm{B}})|^2$, respectively.



%